\documentclass[conference]{IEEEtran}
\usepackage{cite}
\usepackage{amsmath,amssymb,amsfonts}
\usepackage{textcomp}
\usepackage{xcolor}
\usepackage{amsthm}
\def\BibTeX{{\rm B\kern-.05em{\sc i\kern-.025em b}\kern-.08em
    T\kern-.1667em\lower.7ex\hbox{E}\kern-.125emX}}
    
\usepackage{amsmath}
\usepackage{hyperref}
\usepackage{soul}
\usepackage{amssymb}
\usepackage{graphicx}
\usepackage{algorithm}
\usepackage{algpseudocode}
\usepackage{xspace}
\usepackage[caption=false,font=small]{subfig}
\usepackage{booktabs}

\algblockdefx{MRepeat}{EndRepeat}{\textbf{repeat $k$ times}}{}
\algnotext{EndRepeat}

\newcommand{\floor}[1]{\left\lfloor #1 \right\rfloor}
\newcommand{\BCMD}{\texttt{BCMD}\xspace}
\newcommand{\BCMDone}{\texttt{BCMD-M}\xspace}

\newcommand{\BCMDd}{\texttt{BCMD-$\delta$}\xspace}
\newcommand{\setcover}{\texttt{SetCover}\xspace}
\newcommand{\bigO}{\ensuremath{\mathcal{O}}\xspace}
\newcommand{\NP}{\ensuremath{\mathbf{NP}}\xspace}
\newcommand{\poly}{\ensuremath{\mathbf{P}}\xspace}

\newtheorem{cor}{Corollary}
\newtheorem{defi}{Definition}   
\newtheorem{theorem}{Theorem}    
\newtheorem{lemma}{Lemma}     
\newtheorem{proposition}{Proposition}
\newtheorem{prob}{Problem}
    
\begin{document}

\title{Diameter Minimization by Shortcutting with Degree Constraints
\thanks{Both authors are supported by the ERC Advanced Grant REBOUND (834862),
the EC H2020 RIA project SoBigData (871042), and the Wallenberg AI, Autonomous Systems
and Software Program (WASP) funded by the Knut and Alice Wallenberg Foundation}
}

\author{
\IEEEauthorblockN{1\textsuperscript{st} Florian Adriaens}
\IEEEauthorblockA{\textit{KTH Royal Institute of Technology} \\
Stockholm, Sweden \\
adriaens@kth.se}
\and
\IEEEauthorblockN{2\textsuperscript{nd} Aristides Gionis}
\IEEEauthorblockA{\textit{KTH Royal Institute of Technology} \\
Stockholm, Sweden\\
argioni@kth.se}
}

\maketitle

\begin{abstract}
We consider the problem of adding a fixed number of new edges to an undirected graph 
in order to minimize the diameter of the augmented graph,
and under the constraint that the number of edges added for each vertex is bounded by an integer.
The problem is motivated by network-design applications, 
where we want to minimize the worst case communication in the network
without excessively increasing the degree of any single vertex, so as to avoid additional overload.
We present three algorithms for this task, each with their own merits.
The special case of a matching augmentation --when every vertex can be incident to at most one new edge-- is of particular interest, for which
we show an inapproximability result, and provide bounds on the smallest achievable diameter when these edges are added to a path. Finally, we empirically evaluate and compare our algorithms on several real-life networks of varying types.

\end{abstract}

\begin{IEEEkeywords}
approximation algorithms, network design, edge augmentation, diameter reduction.
\end{IEEEkeywords}

\section{Introduction}
The diameter of a graph is defined as the greatest distance between any pair of vertices. It is a fundamental notion of a network, measuring the worst-case point-to-point distance in information networks, social networks, 
and communication networks.
Ensuring a small diameter is a crucial property in network-design applications, e.g.,
minimizing latency in multi\-core processor networks~\cite{benini2002networks}, or 
forming a small-world network to maximize the influence of a campaign~\cite{laoutaris2008bounded}.

There has been a considerable amount of research on the problem of augmenting 
undirected graphs with new edges in order to minimize the diameter of the resulting graph. 
This operation has been described in the literature as \emph{short\-cutting}~%
\cite{DemaineShortcut, Meyerson, TAN201791, Chepoi};
and the newly-added edges are referred to as \emph{short\-cut edges}.
Li, McCormick and Simchi-Levi \cite{LiDiam} introduced the \emph{Bounded Cardinality Minimum Diameter} (\BCMD) problem, 
where the goal is to add at most $k$ short\-cut edges 
so as to minimize the diameter of the augmented graph.

\begin{figure*}[t]
\centering
\subfloat[HIV network.\label{fig:introdiam}]{\includegraphics[width=0.6\columnwidth,trim={0.5 1.3cm 0.5cm 1cm},clip]{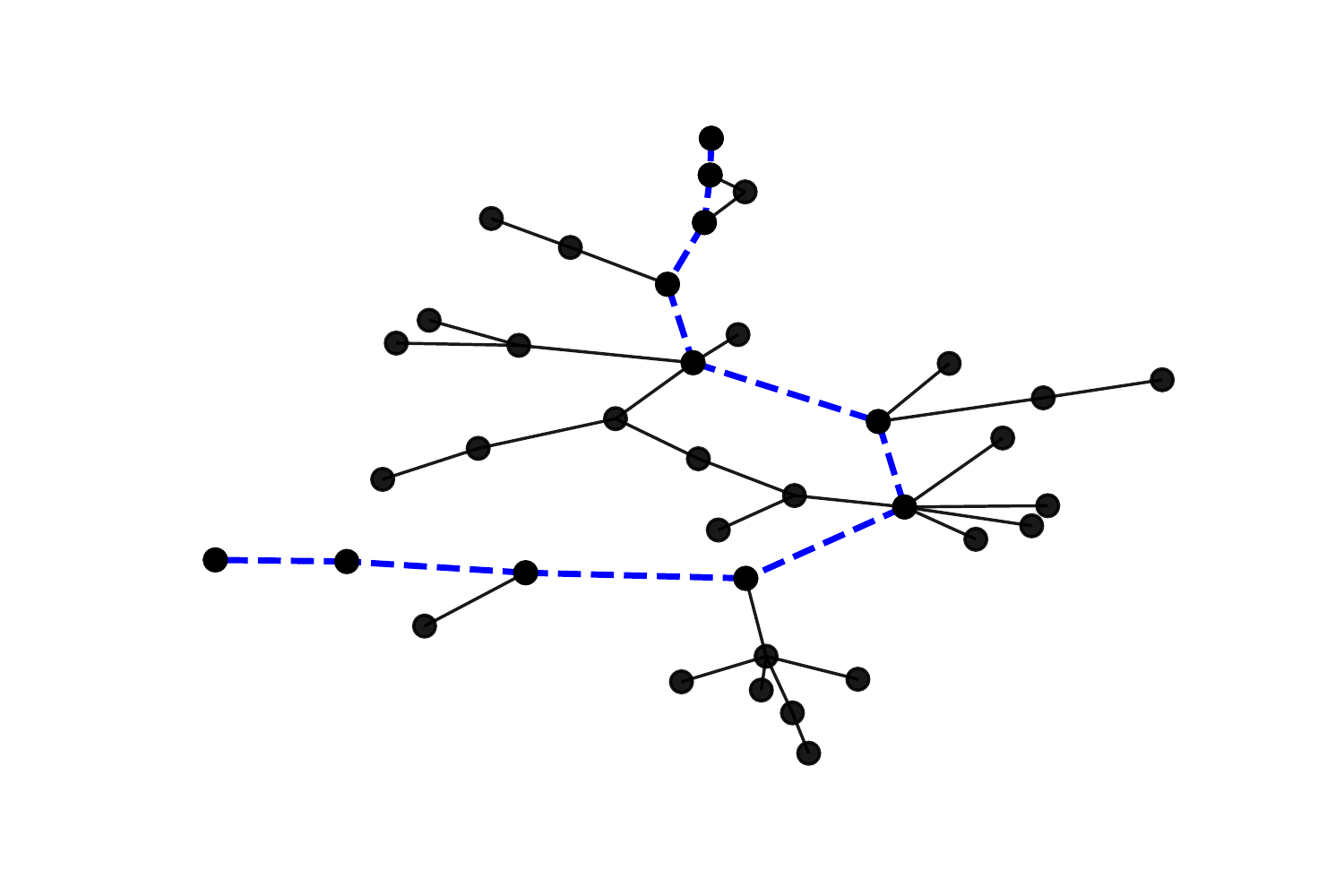}}
\subfloat[degree budget $\delta=3$.\label{fig:introa}]{\includegraphics[width=0.6\columnwidth,trim={0.5 1.3cm 0.5cm 1cm},clip]{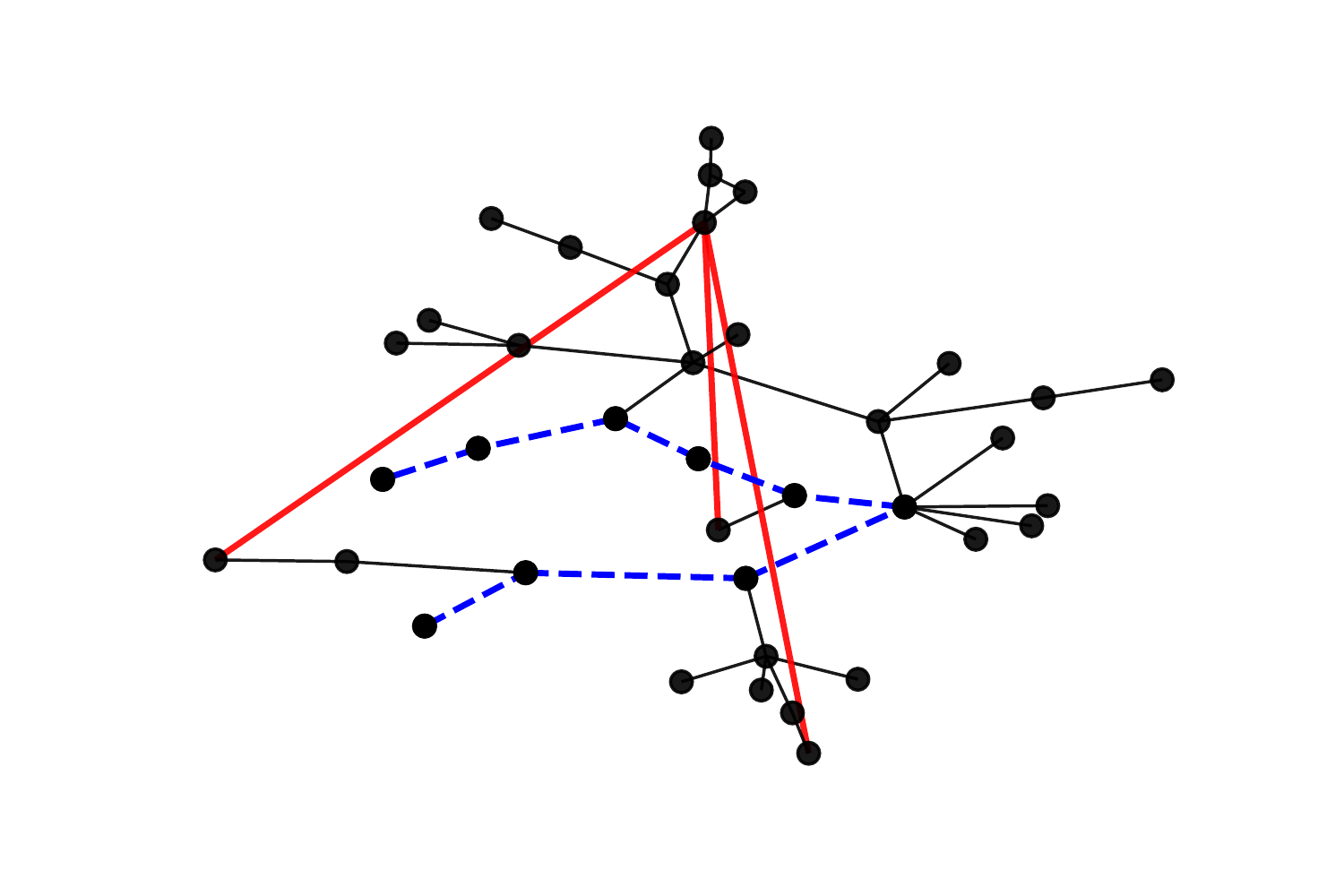}}
\subfloat[degree budget $\delta=1$.\label{fig:introb}]{\includegraphics[width=0.6\columnwidth,trim={0.5 1.3cm 0.5cm 1cm},clip]{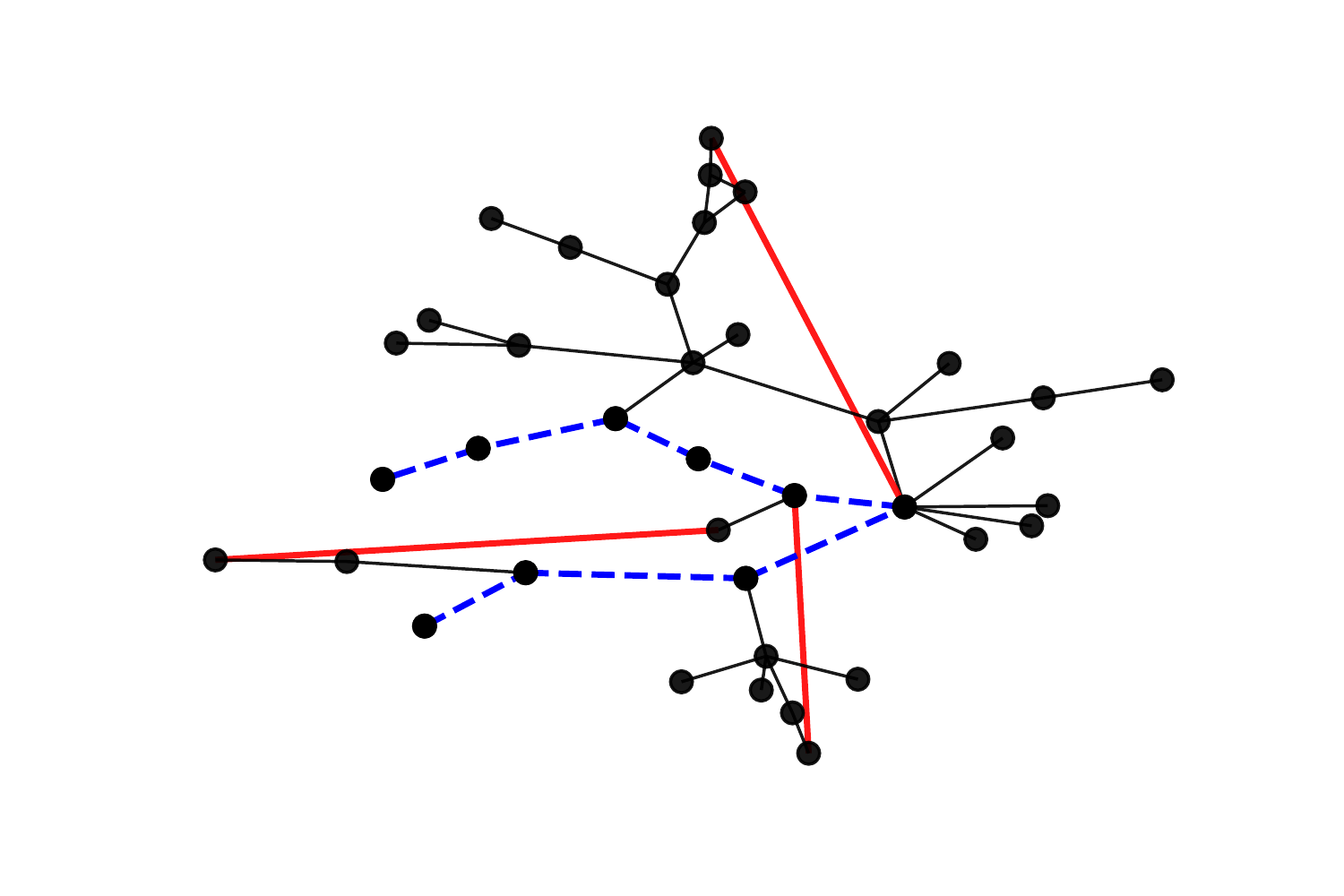}}
\caption{The result of our algorithm (Section~\ref{sec:smallk}) after adding $k=3$ shortcut edges (indicated by the red edges) to the HIV network \cite{Auerbach1984ClusterOC,Konect}. Fig. (a) shows the original network. In (b) we have shortcut the network with a degree increase constraint of $\delta=3$ for each vertex. In (c) we set $\delta=1$, so the augmentation has to form a matching. The diameter of the original network is 10, as indicated by a longest shortest path (blue dashed edges). In both (b) and (c), the diameter of the augmented network is reduced to 8. 
\label{fig:introhiv}}
\end{figure*}

A potential downside of the \BCMD\ problem formulation is that optimal solutions (and their approximations) 
might increase the degree of a single vertex by a substantial amount.
In fact, this is a guaranteed side effect for solutions obtained by the known approximation algorithms for \BCMD (see Section~\ref{sec:rw}), 
as a common theme in many of these algorithms is that they first partition the graph into $k+1$ non-overlapping parts (the clustering step), 
and then connect the clusters by picking one cluster center and connecting it to the centers of all other clusters. The latter step is called the \emph{star-short\-cutting} step and adds at most $k$ short\-cut edges to the graph, by increasing the degree of a single vertex by $k$.
For certain applications, however, it is desirable to limit the increase of the degree of any single vertex.
Due to physical, economical, or other limitations, many real-life network problems introduce 
such a degree constraint \cite{ChungAlt, TAN201791}.
For example, the work of Bokhari and Raza \cite{Raza} was motivated by a question on how to decrease the diameter of a computer network, by adding additional 
links, under the constraint that no more than one I/O port is added to each~processor.
A second example is in the area of social networks:
a service provider might be interested in recommending new friendships to users 
with the aim of increasing the overall connectivity of the network, 
so as to facilitate information diffusion and 
reducing polarization~\cite{garimella2017reducing,haddadan2021repbublik,Rubenpolarization}. In \cite{Rubenpolarization} the largest distance between two members of different groups (i.e., the \emph{colored} diameter) has been used a measure of polarization.
Recommending too many friendships to one individual user might not result in many actual links being materialized, due to the risk of overburdening that user.
A better strategy might be to limit the number of recommendations per individual~user.

Motivated by these application scenarios and 
the need for a different algorithmic approach than prior algorithms for \BCMD{}, 
as \emph{star-short\-cutting} is not allowed anymore,
we introduce and study a degree-constrained generalization of \BCMD{} (see Problem~\ref{defi:prob}). 
In this variant, denoted as \BCMDd{}, the increase of the degree of each vertex is limited to at most $\delta$. The original \BCMD{} problem corresponds to \BCMDd{} with $\delta=k$.
Studying the \BCMD{} problem with degree constraints was left as an open question in the recent work of Tan, E. J. Van Leeuwen and J. Van Leeuwen.~\cite{TAN201791}. Fig.~\ref{fig:introhiv} shows an illustrative example of the \BCMDd{} problem, for different settings of $\delta$. In Fig.~\ref{fig:introa} we have an unlimited degree increase budget, corresponding to the original \BCMD{} problem. 
In Fig.~\ref{fig:introb} we illustrate a solution to the \BCMDd{} problem with $\delta=1$:
the vertex degrees are allowed to increase by at most~one, meaning that the newly added edges have to form a matching in the augmented graph.

\smallskip
\noindent
{\bf Preliminaries.}
All graphs $G=(V,E)$ in this paper are simple, undirected and unweighted.
Let $n=|V|$, and $m=|E|$.
The distance $d_G(u,v)$ between two vertices $u,v \in V$ is defined as the number of edges in a shortest path between $u$ and $v$ ($\infty$ if no connecting path exists).
The eccentricity $e_G(u) = \max_{v \in V} d_G(u,v)$ of a vertex $u$ is the largest distance between $u$ and any other vertex. The diameter $D_G = \max_{u,v \in V}d_G(u,v)$ is the maximum eccentricity of any vertex. 
This distance between two sets of vertices is always defined as the shortest distance between any two vertices, one from each set.
For $X \subseteq V$, the induced subgraph $G[X]$ of $G$ by $X$ is the graph whose vertex set is $X$, and whose edge set consists of all of the edges in $E$ that have both endpoints in $X$.
In case we omit the subscript in a notation, we always refer to the original graph $G$.

\smallskip
\noindent
{\bf Results and outline.}
%
Problem~\ref{defi:prob} introduces a generalization of the \BCMD{} problem 
that restricts the maximum degree increase of each vertex.
Specifically, given a budget $\delta$ for each vertex,
we ask to augment the graph by adding at most  $k$ new edges
so as to minimize the diameter of the augmented graph, 
and the degree of each vertex increases by at most $\delta$.

\begin{prob}[Bounded Cardinality Minimum Diameter with degree constraints (\BCMDd{})\label{defi:prob}]
Given a graph $G=(V,E)$ and integers $k,\delta \geq 1$, 
find a set $M$ of at most $k$ non-edges in $G$ that minimize 
the diameter of the augmented graph $G' = (V,E \cup M)$, 
and for all $v \in V$ it holds that $\text{degree}_{G'}(v) \leq \text{degree}_{G}(v)+\delta$. 
\end{prob}

In the most restrictive setting, which corresponds to $\delta=1$, 
each vertex can be incident to at most one shortcut edge, 
so the augmentation has to form a matching.
Since this problem is of particular interest, we denote it as \BCMDone{} (Bounded Cardinality Minimum Diameter with Matching constraint).

Our results are summarized as follows:

\begin{enumerate}
\item In Section~\ref{sec:path} we present lower and upper bounds on the optimum value of the \BCMDone{} problem when the input graph is 
an $n$-vertex path, 
thereby extending a result of Chung and Garey~\cite{ChungAlt} to the setting of a matching augmentation.
\item We give three algorithms for the \BCMDd{} problem in Section~\ref{sec:gengraphs}. Section~\ref{sec:logkapp} details a $\bigO(\log_{\delta+1} k)$-approximation algorithm for connected graphs. In Section~\ref{sec:smallk} we give a constant-factor approximation 
in the case that $k \leq \sqrt{\delta n}-1$. Section~\ref{sec:heuristic} details an intuitive heuristic without guarantees. All algorithms need $\bigO(km)$ time, and hence scale quite well for limited values of $k$.

\item In Section~\ref{sec:inapprox} we show that there exists no 
$(\frac{4}{3}-\epsilon)$-approximation  for \BCMDone{}, assuming $\poly \neq \NP$.
\item In Section~\ref{sec:exps} we empirically evaluate and compare the performance of our algorithms proposed in Section~\ref{sec:gengraphs} on real-life networks of varying types and sizes.
\end{enumerate} 

\section{Related work}
\label{sec:rw}
Several constant-factor approximation algorithms for the \BCMD problem, and weighted variants thereof, are known.
Besides proving \NP-completeness,
Li, McCormick and Simchi-Levi \cite{LiDiam} proposed an elegant $(4+\frac{2}{D^*})$-approximation algorithm for \BCMD, 
where $D^*$ denotes the smallest achievable diameter.
Bil\`{o}, Gual\`{a} and Proietti \cite{BILO201212} refined the performance analysis of \cite{LiDiam}, 
and showed that the algorithm in fact guarantees a $(2+\frac{2}{D^*})$-approximation.
Demaine and Zadimoghaddam \cite{DemaineShortcut} considered weighted graphs, 
with the requirement that all short\-cut edges have an equal nonnegative weight, 
and
they provided a $(4+\epsilon)$-approximation for arbitrary $\epsilon>0$.
Dodis and Khanna \cite{Dodis} generalized the problem by assigning costs to the short\-cut edges, 
and requiring that the sum of the costs of all short\-cut edges does not exceed a given budget parameter.
They showed that the general variant of \BCMD{}, 
with arbitrary positive edge weights and arbitrary nonnegative short\-cut edge costs, 
admits a $(2+\frac{2}{D^*})$-approximation if one is allowed to exceed the budget by a factor of $\bigO(\log k)$.
Finally, Frati et al. \cite{Frati} showed that the general variant of \BCMD{} 
admits a fixed-parameter tractable 4-approximation algorithm.

\smallskip
\noindent
{\bf Prior work on short\-cutting with degree constraints.}
%
The problem of short\-cutting a graph while respecting degree constraints has been studied before.
Chung and Garey \cite{ChungAlt} provided lower and upper bounds on the smallest achievable diameter when 
edges are added to a path.
They posed it as an open question to extend their results 
when there is a constraint on the maximum degree increase of each vertex.
In Section~\ref{sec:path} we extend their bounds to the case of a matching augmentation ($\delta=1$).
Bollob\'{a}s and Chung \cite{BollobasRandom} showed that adding a random matching of maximum possible size to a cycle 
gives a graph with diameter close to the optimum value, which is about $\log_2 n$.
Bokhari and Raza \cite{Raza} proved that by adding at most $\floor{\frac{n}{2}}$ matching edges, 
any connected graph can be short\-cut to a diameter of $\bigO(\log n)$.
More recently, Tan et al. \cite{TAN201791} improved the results of \cite{Raza} and showed that  
$\bigO(\frac{n}{\log n})$ matching edges are sufficient to achieve a diameter of $\bigO(\log n)$.
Moreover, they provided several complexity results on the dual problem (see also \cite{Frati,GAO20131626}),
i.e., 
minimizing the number of edges 
that need to be added to achieve a target diameter, while respecting degree constraints.

\section{Warm up: Shortcutting a path with matching edges}
\label{sec:path}

We present several results on the \BCMDone{} problem when the input graph is an $n$-vertex path.
The ideas presented here are intuitive and they serve as a building block for our shortcutting algorithms on general graphs in Section~\ref{sec:gengraphs}.

Chung and Garey~\cite{ChungAlt} provided lower and upper bounds on the smallest achievable diameter 
when~$k$ short\-cut edges are added to an $n$-vertex path, without any constraints on the maximum increase of the degrees.
They proved that it is not possible to achieve a diameter smaller than $\frac{n}{k+1}-1$, 
and gave an algorithm that reduces the diameter to $\frac{n}{k+1}+3$.
In the following sections we extend their bounds on the smallest achievable diameter 
when the vertex-degree increase is limited to~one, i.e., when $k$ matching edges are added to a path.

\subsection{Lower bound}
\label{sec:low}
Let $M(n,k)$ denote the smallest achievable diameter after adding $k$ matching edges to an $n$-vertex path.
The lower bound of Chung and Garey~\cite{ChungAlt} also applies to our setting, 
and we immediately find that $M(n,k) \geq \frac{n}{k+1}-1$.
On the other hand, it is known that any graph with maximum degree three has diameter at least $\log_2(n)-2$
\cite{BollobasRandom}.
Theorem~\ref{thm:lbpath} unifies both results into one lower bound.
Theorem~\ref{thm:ubpath} gives a method that achieves this lower bound, 
up to constant factors.

%
%
%
%

\begin{theorem}
\label{thm:lbpath}
For all $1 \leq k \leq n/2$, 
the smallest achievable diameter after adding $k$ matching edges to an $n$-vertex path satisfies
\begin{align}\label{eq:lb} 
M(n,k) \geq \frac{n}{2(k+1)}+\log_2(k+1)-2.
\end{align}
\end{theorem}

\begin{IEEEproof}
Consider the rooted shortest-path tree $T$ formed by a breadth-first-search (BFS) in the optimal augmentation (i.e., a short\-cutting that achieves diameter $M(n,k)$), starting from a vertex $v$, until we encounter all vertices.
The height $h_T$ of $T$ 
is a lower bound for $M(n,k)$. 
We provide a lower bound to $h_T$ to establish the result.
The height $h_T$ is minimized 
when the degree-three vertices are explored first, since the degree-two vertices do not lead to any branching in $T$.
So we may assume that the first levels of $T$ are filled with degree-three vertices (in particular, the root vertex $v$ has degree three), after that the degree-two and vertices, and finally the degree-one vertices.
Furthermore, we may assume that none of the added $k$ short\-cut edges are incident to 
any of the two endpoints of the path. Since in this case, we have the maximum possible number of degree-three vertices, which yields the smallest possible height $h_T$. 
After the addition of any such set of $k$ matching edges, there are $2k$ vertices of degree three, $n-2k-2$ vertices of degree two, and the two endpoints of the path have degree one.


Let level $i$ be the largest level containing degree-three vertices, 
where we place the root vertex $v$ at level~0.
There are $2k$ degree-three vertices, so it follows that $i = \left \lceil{\log_2(\frac{2k+2}{3})}\right \rceil\geq \log_2(k+1)-1$.
After encountering the degree-three vertices in $T$, we encounter the degree-two vertices and ultimately the two degree-one endpoints of the path.
Some of the degree-two vertices might be at level $i$, but from level $i+1$ the tree consists only of vertices with degree at most two.
This implies that the tree $T$ will not branch any further after level $i+1$.
We now quantify how many additional levels the remaining $n-2k$ vertices of degree at most two induce.
Note that there are at most $k+1$ degree-three vertices at level $i$, and this is tight if the level $i$ is filled fully with degree-three vertices.
It follows that the additional number of levels induced by the $n-2k$ remaining vertices of at most degree-two is at least $\frac{n-2k}{2(k+1)}$, since the degree-three vertices at level $i$ might induce two branches of degree-two vertices in level $i+1$.
Hence,
\begin{align*}
M(n,k) \geq h_T &\geq \frac{n-2k}{2(k+1)} + \log_2(k+1)-1 \\
&\geq \frac{n}{2(k+1)} + \log_2(k+1)-2.
\end{align*}

%
%
\end{IEEEproof}

\subsection{Upper bound}
\label{sec:lowup}

Next we describe a 
procedure that achieves a matching upper bound, up to constant factors.
\begin{defi}
\label{defi:mooretree}
A full (or Moore) d-tree is an (undirected) rooted tree in which every vertex has degree $d$, and all levels are filled to maximum size except possibly the leaf level \cite[Section 2.1.2]{TAN201791}.
\end{defi}


\begin{theorem}\label{thm:ubpath}
For all $1\leq k \leq n/2$, 
the smallest achievable diameter after adding $k$ matching edges to an $n$-vertex path satisfies
\begin{align}\label{eq:ub}
M(n,k) \leq \frac{n}{k+1} + 4\log_2(k+1)+ 1.
\end{align}
\end{theorem}

\begin{IEEEproof}
First, we discuss a procedure when $k \leq \frac{n}{3}-1$.
We divide the path into $k+1$ non-overlapping intervals of roughly equal size. For every interval we find the center vertex, i.e., the midpoint of the interval. If there is a tie between two centers, choose one arbitrarily. Each interval will contain at least $\left \lfloor{n/(k+1)}\right \rfloor$ vertices and at most $\left \lceil{n/(k+1)}\right \rceil$ vertices. The distance from any vertex to the center of its respective interval is at most $\left \lceil{n/(k+1)}\right \rceil/2$.
Since $k \leq \frac{n}{3}-1$, every interval has at least three vertices.
In each interval we select a line \emph{segment} (a path of length two) consisting of the left-neighbor of the center, the center itself, and the right-neighbor of the center (see Fig.~\ref{fig:fta}).
We now connect the segments by embedding them into a \emph{full 3-tree} (see Definition~\ref{defi:mooretree}). The segments are regarded as supernodes in the tree.
Take an arbitrary segment as the root in the tree, 
and short\-cut each vertex in the segment to the centers of three other segments. 
Each of these three segments will now be short\-cut to the centers of two new segments and so on, 
until all segments are part of the tree. Note that all vertices are incident to at most one short\-cut edge.
Fig.~\ref{fig:ftb}) gives an example of how such a construction looks like.

\begin{figure}[t]
\centering
\subfloat[\label{fig:fta}]{\scalebox{0.5}{\includegraphics[trim={0 0.4cm 0.1cm 0},width=1\linewidth]{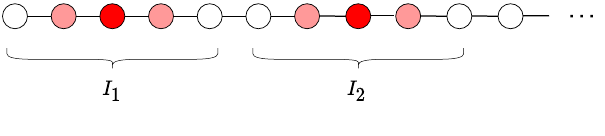}}}
\subfloat[\label{fig:ftb}]{\scalebox{0.5}{\includegraphics[width=1\linewidth]{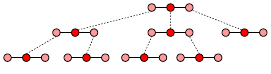}}}

\caption{a) Dividing the path into $k+1$ intervals $I_i$. A segment consists of an interval center (red) together with its two neighbors (pastel red). b) Embedding the segments into a full 3-tree. Dashed edges indicate the newly added edges.\label{fig:fulltree}}
\end{figure}

The height of this tree is at most $\log_2(k+1)$, and it takes at most two steps to traverse each segment.
So the distance between any two center vertices in the short\-cutted graph is at most $4\log_2(k+1)$.
The distance from an arbitrary vertex to the closest segment in the same interval is at most $\left \lceil{n/(k+1)}\right \rceil/2$.
The total diameter D of the short\-cutted graph is hence bounded by
\begin{align*}
D \leq \left \lceil\frac{n}{k+1}\right \rceil + 4\log_2(k+1) \leq {\frac{n}{k+1}} + 4\log_2(k+1)+ 1.
\end{align*}

In case that $k > \frac{n}{3}-1$, it suffices to add $\frac{n}{3}-1$ edges in the aforementioned manner.
Indeed, a straightforward calculation shows that the bound (\ref{eq:ub}) is increasing as a function of $k$, when $k > \frac{n \ln(2)}{4}-1$.
\end{IEEEproof}


\begin{figure}[t]
\centering
\scalebox{0.65}{\includegraphics[width=1\linewidth]{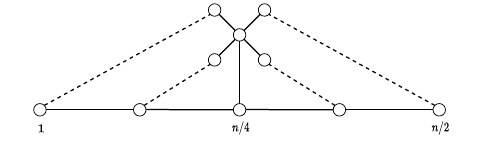}}
\caption{A graph with diameter $n/2-1$ that is short\-cut to a graph with diameter four by adding $k=n/2-1$ matching short\-cut edges (dashed edges).\label{fig:toygraph}}
\end{figure}

\subsection{Achievable diameter in connected graphs}
\label{sec:achievablediam}
Chung and Garey~\cite[Section 3]{ChungAlt} used their lower bound on the achievable diameter of a short\-cutted path to show that any connected graph with diameter $D$ cannot be short\-cut to a diameter less than $\frac{D+1}{k+1}-1$, using at most $k$ edges. 
One can try to extend their result to our case of matching short\-cut edges, but the example in Fig.~\ref{fig:toygraph} shows that their bound is tight in certain cases.
The example shows a graph with diameter $n/2-1$ (assume that $n$ is even) 
that is being short\-cut to a graph with diameter four
by adding $k=n/2-1$ matching edges. 
This is in contrast with short\-cutting a path, which also has diameter linear in $n$. However, unlike the example in Fig.~\ref{fig:toygraph}, a path has constant vertex degrees and this restricts the achievable diameter. Indeed, Theorem~\ref{thm:lbpath} states that adding any $k=n/2-1$ matching edges to an $n$-vertex path gives a graph with diameter at least $\log_2(n)-2$.
For short\-cutting general connected graphs with diameter $D$ with matching edges, it appears one cannot improve the lower bound of $\bigO(\frac{D}{k})$ as given by \cite{ChungAlt} without taking into account additional information about the graph (e.g., degrees).

\section{Shortcutting general graphs}
\label{sec:gengraphs}
We present three algorithms for shortcutting general graphs with degree constraints.
In Section~\ref{sec:logkapp} we present an algorithm that achieves a $\bigO(\log_{\delta+1} k)$-approximation 
for connected graphs.
In Section~\ref{sec:smallk} we present a constant-factor approximation when $k \leq \sqrt{\delta n}-1$.
In Section~\ref{sec:heuristic} we detail a fast heuristic without approximation guarantees.
All three algorithms need $\bigO(k m)$ time, 
making them scalable on most real-life networks when $k$ is not too~large.


\subsection{A $\bigO(\log_{\delta+1} k)$-approximation for connected graphs.}
\label{sec:logkapp}

The high-level strategy is reminiscent of the path case.
After finding at most $k+1$ small connected segments in the graph, we embed them into a full tree by connecting the segments using at most $k$ shortcut edges. The branching factor of the tree will depend on the degree budget $\delta$ and the size of the segments.
The main difference with the path case is that we will require the segments to be far apart (this will later be defined more precisely).
Our strategy consists of first finding a maximal family of vertex-disjoint segments. 
From this family, we select a set of at most $k+1$ segments that are far apart, 
using a similar idea as the $k$-center heuristic of Dyer and Frieze \cite{DYER1985285,gonzalez1985clustering}.
This heuristic has been used in previous algorithms for
the \BCMD{} problem~\cite{BILO201212,Frati,LiDiam}, and allows us to bound the maximum distance, in terms of the optimum diameter, of a vertex to the chosen set of $k+1$ segments (Corollary~\ref{cor:dtotal}).

We start by formally defining a segment.

\begin{defi}
A $\beta$-\emph{segment} $C$, for $1 \leq \beta \leq n$, is a set of vertices $C \subseteq V$ with $|C| = \beta$, such that the induced subgraph $G[C]$ is connected.
\end{defi}

The following proposition bounds the distance between a vertex and a maximal family of vertex-disjoint segments.

\begin{proposition}
\label{prop:maxfam}
Assume $G$ is connected. For any \emph{maximal} family $\mathcal{C} = \{C_1, \ldots, C_{\alpha}\}$ of vertex-disjoint $\beta$-segments, it holds that $\forall v \notin \cup_{i=1}^{\alpha} C_i$: $d_G(v, \mathcal{C}) \leq \beta-1$.
\end{proposition}
\begin{IEEEproof}
Suppose for some $v \notin \cup_{i=1}^{\alpha} C_i$ we have $d_G(v, \mathcal{C}) > \beta-1$.
Since $G$ is connected, there is a shortest path from $v$ to $\cup_{i=1}^{\alpha} C_i$ that consists of at least $\beta$ vertices that are not in $\cup_{i=1}^{\alpha} C_i$, contradicting the maximality of $\mathcal{C}$. 
\end{IEEEproof} 




\begin{algorithm}[t]
\caption{Log. approximation \BCMDd{}.}\label{algo:segmentcenter}
\begin{algorithmic}[1]
\Require graph $G$, $k \geq 1$, $\delta \geq 1$ and $3 \leq \beta \leq n$. 
\State $\mathcal{X} \leftarrow \emptyset$
\State $\mathcal{C}$ = a maximal family  of disjoint $\beta$-segments.

\For{$\min\{k+1,|\mathcal{C}|\}$ times}
\State Pick segment $C^* \in \mathcal{C}\setminus \mathcal{X}$ with $d_G(C^*, \mathcal{X}) = \max_{C \in \mathcal{C}\setminus \mathcal{X}} d_G(C,\mathcal{X})$.
\State $\mathcal{X} \leftarrow \mathcal{X} \cup \{C^*\}$.
\EndFor

\State Connect the segments in $\mathcal{X}$ into a full $d$-tree with $d = \beta \, \delta$, such that every vertex has at most $\delta$ incident shortcut edges. See Fig.~\ref{fig:ftb} for an example with $\delta=1$ and $\beta=3$.
\end{algorithmic}
\end{algorithm}

Let $D_k^*$ denote the minimum achievable diameter after adding at most $k$ edges to $G$ (the optimum of \BCMD{}), and let  $D^*_{k,\delta}$ be the minimum achievable diameter 
after adding at most $k$ edges to $G$, without increasing the degree of any vertex by more than $\delta \geq 1$ (the optimum of \BCMDd{}).

After running Algorithm~\ref{algo:segmentcenter} (lines 1--6), we prove (Lemma~\ref{lem:dkopt}) that the distance from any remaining segments in $\mathcal{C}\setminus \mathcal{X}$ to $\mathcal{X}$ is bounded by $D^*_{k,\delta}$.
The first segment in $\mathcal{X}$ can be chosen arbitrarily. 
Algorithm~\ref{algo:segmentcenter} (lines 3--6) then selects segments with the largest distance from the already selected ones. 
\begin{lemma}
\label{lem:dkopt}
For all  $\beta$-{segments} $C \in \mathcal{C}\setminus \mathcal{X}$ it holds $d_G(C, \mathcal{X}) \leq D^*_{k,\delta}$.
\end{lemma}

\begin{IEEEproof}
We can assume that $\mathcal{C}\setminus \mathcal{X} \neq \emptyset$.
This implies that $k+1<|\mathcal{C}|$, and thus $|\mathcal{X}| = k+1$.
Suppose there exists a $\beta$-{segment} $C \in \mathcal{C}\setminus \mathcal{X}$ such that $d_G(C, \mathcal{X}) > D^*_{k}$.
Algorithm~\ref{algo:segmentcenter} greedily selects the segments that are furthest away 
from the already selected set of segments, 
This implies that when the iteration (lines 3--6) terminates
we have a set of $|\mathcal{X}|+1 = k+2$ segments with pairwise distances between them 
that is larger than $D^*_{k}$.
In particular, we have a set of at least $k+2$ vertices with pairwise distances larger than $D^*_{k}$.
Now we use the following proposition 
--- the proof can be found in the paper of Frati et al.~\cite[Claim 1]{Frati}:
\begin{proposition}
\label{propFrati}
\cite[Claim 1]{Frati} Let $X \subseteq V$ be a set of vertices with $|X| \geq 3$ 
and $d(u,v) > \alpha$, for all $u \neq v \in X$, for some $\alpha>0$.
Then for any graph $G'=G\cup\{e\}$, 
where $e \in {V\choose2}$, 
there exists a subset $X' \subseteq X$ with $|X'|=X-1$, such that $d_{G'}(u,v) > \alpha$, 
for all $u \neq v \in X'$.
\end{proposition}
By iteratively applying Proposition~\ref{propFrati}, 
any graph $G'$ obtained by adding any $k$ shortcut edges (not necessarily with degree-constraints) to $G$ 
will have a set of at least two vertices with pairwise distances in $G'$ strictly larger than $D^*_{k}$, 
which contradicts the optimality of $D^*_k$.
So it must be that $\forall C \in \mathcal{C}\setminus \mathcal{X}$ it holds that $d_G(C, \mathcal{X}) \leq D^*_{k}$, and the result follows since $D_k^* \leq D^*_{k,\delta}$.
\end{IEEEproof}


Lemma~\ref{lem:dkopt} enables us to bound the distance from any vertex $v$ to the set of segments~$\mathcal{X}$, as given by the following corollary.

\begin{cor}
\label{cor:dtotal}
Assume $G$ is connected. For all vertices $v \in V$ it holds $d_G(v,\mathcal{X}) \leq \beta-1+D^*_{k,\delta}$.
\end{cor}
\begin{IEEEproof}
If $|\mathcal{C}| \leq k+1$, then $d_G(v,\mathcal{X}) \leq \beta-1$ since $\mathcal{X} = \mathcal{C}$ after running Algorithm~\ref{algo:segmentcenter} (lines 1--6) and because $\mathcal{C}$ is a maximal family of $\beta$-{segments} (Proposition~\ref{prop:maxfam}).
If $|\mathcal{C}| > k+1$, then $\mathcal{C}\setminus \mathcal{X} \neq \emptyset$ 
and the result follows from Lemma~\ref{lem:dkopt} and the triangle inequality. 
\end{IEEEproof}




Now we are ready to analyse Algorithm~\ref{algo:segmentcenter}:

\begin{theorem}
\label{thm:appratio}
Assume $G$ is connected. Algorithm~\ref{algo:segmentcenter} returns a shortcutted graph whose diameter is  
at most $2(\beta-1+D^*_{k,\delta}+\beta\log_{\beta\delta-1}(k+1))$, by adding at most $k$ edges to $G$.
\end{theorem}

\begin{IEEEproof}
By Corollary~\ref{cor:dtotal}, the distance in $G$ (and thus, also in the augmented graph) of any vertex to its nearest segment in $\mathcal{X}$ is at most $\beta-1+D^*_{k,\delta}$.
The constructed full $d$-tree has height at most $\log_{\beta\delta-1}$ (since $\beta \geq 3$ and $\delta \geq 1$, otherwise there is no branching of the tree), and it takes at most $\beta-1$ steps to traverse each $\beta$-segment.
Hence, the total distance in the augmented graph between any two segments in $\mathcal{X}$ is at most $2\beta\log_{\beta\delta-1}|\mathcal{X}|$.
So the total diameter is at most $2(\beta-1+D^*_{k,\delta}+\beta\log_{\beta\delta-1}|\mathcal{X}|)$.
The result follows since $|\mathcal{X}|\leq k+1$.

Note that Algorithm~\ref{algo:segmentcenter} (line 7) adds at most $k$ shortcut edges to the original graph; 
constructing the full $d$-tree needs at most $|\mathcal{X}|-1$ edges, and 
$|\mathcal{X}|\leq k+1$ (line 3).
\end{IEEEproof}

\begin{cor}
\label{cor:appratio}
Assume $G$ is connected. Algorithm~\ref{algo:segmentcenter} is a $\bigO(\log_{\delta+1} k)$-approximation for the \BCMDd{} problem,  
for the choice $\beta = 3$.
\end{cor}

\begin{IEEEproof}
If $D^*_{k,\delta} \geq \log_{3\delta-1}(k+1)$, then Theorem~\ref{thm:appratio} gives a constant-factor guarantee. 
If $D^*_{k,\delta} < \log_{3\delta-1}(k+1)$ then the resulting diameter guarantee is $\bigO(\log_{\delta+1} k)$, 
and the result follows since $D^*_{k,\delta} \geq 1$.
\end{IEEEproof}


\smallskip
\noindent
{\bf Running time.} Algorithm~\ref{algo:segmentcenter} needs $\bigO(km)$ time for connected graphs when $\beta=3$. Finding a maximal family of 3-segments (line 2) can be done in linear time $\bigO(m)$ (see e.g., \cite[Section 3.1]{TAN201791}).
Lines 3--5 can be implemented in $\bigO(km)$ time (similarly as in the k-center heuristic \cite{DYER1985285}); in every iteration, keep a dictionary of the distances from all segments to the current set $\mathcal{X}$.
After adding $C^*$ to $\mathcal{X}$, we update the dictionary efficiently by first computing the distances from $C^*$ to all other segments (by doing a BFS in $\bigO(m)$ time), and then for each segment taking the minimum of the old dictionary value and this newly computed distance.


\subsection{Constant-factor approximation for small $k$}
\label{sec:smallk}

\begin{algorithm}[t]
\caption{Constant approximation \BCMDd{}.}\label{algo:constalgo}
\begin{algorithmic}[1]
\Require graph $G$, $1\leq k \leq \sqrt{\delta n}-1$, $\delta \geq 1$. 
\State $\{c_i\} = k+1 \text{ cluster centers computed according to \cite{DYER1985285}}.$
\State $C_{\max} \text{ } (c_{\max}) \leftarrow \text{ largest cluster (center)}$.
\State $\forall c_i \neq c_{\max}$: \text{ add edge } $(c_i,v)$ for some $v \in C_{\max}$ \text{ s.t. $v$ is incident to less than $\delta$ new shortcut edges}.
\end{algorithmic}
\end{algorithm}

\begin{figure}[t]
\centering
\scalebox{0.72}{\includegraphics[trim={0 0 1cm 0},clip,width=1\linewidth]{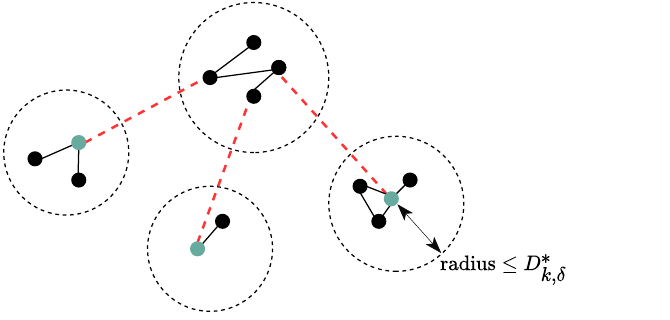}}
\caption{Schematic idea of the constant-factor approximation. In this example, we consider a matching augmentation ($\delta=1$). The radius of each cluster is upper bounded by the optimum diameter $D^*_{k,\delta}$, and the largest cluster has enough vertices to shortcut (dashed red edges) to the centers (green) of the other clusters.\label{fig:sketch}}
\end{figure}

In case when $k \leq \sqrt{\delta n}-1$, we obtain a constant-factor approximation for the \BCMDd{} problem, even for disconnected graphs, using a slightly different approach.
Algorithm~\ref{algo:constalgo} first computes $k+1$ non-necessarily distinct cluster centers (line 1), again according to the $k$-center heuristic of Dyer and Frieze \cite{DYER1985285,gonzalez1985clustering}.
In particular, 
the clustering phase (line 1) computes a set of $k+1$ non-necessarily distinct cluster centers 
$C = \{c_1,\ldots,c_{k+1}\} \subseteq V$ as follows:
the first center $c_1$ is an arbitrary vertex. 
For $2\leq i \leq k+1$, the center $c_i$ is chosen as a vertex that maximizes 
the distance $d_G(c_i,\{c_1,\ldots,c_{i-1}\})$.
The cluster partitions are defined by assigning every non-center vertex to the nearest center, breaking ties arbitrarily.
If $k$ is small (or $\delta$ is large), we can prove that the largest cluster contains enough vertices to shortcut to all other centers, leading to a constant-factor approximation.
Fig.~\ref{fig:sketch} shows an example in case of $\delta=1$.
Lemma~\ref{lem:approxsmallk} formalizes this.

%

\begin{lemma}
\label{lem:approxsmallk}
If $k \leq \sqrt{\delta n}-1$, 
Algorithm~\ref{algo:constalgo} is a $(4+2/D^*_{k,\delta})$-approximation algorithm for \BCMDd{}.
\end{lemma}

\begin{IEEEproof}
The average number of vertices per cluster is $\frac{n}{k+1}$.
So the largest cluster has the capacity to shortcut to at least $\delta\frac{n}{k+1} \geq k$ other vertices.
Thus, Algorithm~\ref{algo:constalgo} (line 3) shortcuts every center from the $k$ other clusters to an arbitrary vertex in the largest cluster, while respecting the degree constraints of all the vertices in the largest cluster. 
In previous work \cite{BILO201212, Frati} 
it was proven that the heuristic \cite{DYER1985285} computes a set of centers $C$ 
such that for all $v \in V$ it holds $d_G(v,C) \leq D_k^*$.
Since $D_k^* \leq D^*_{k,\delta}$, it holds that the radius of each cluster is also upper bounded by $D^*_{k,\delta}$.
So the distance to travel the largest cluster is at most $2D^*_{k,\delta}$, and the diameter of the shortcutted graph is at most $4D^*_{k,\delta}+2$.
\end{IEEEproof}

\smallskip
\noindent
{\bf Running time.}
The running time of Algorithm~\ref{algo:constalgo} is $\bigO(km)$, since the bottleneck is the $k$-center heuristic from \cite{DYER1985285}.

\medskip
\noindent
{\bf Extending to larger depths.}
\label{sec:extending}
We can strengthen the idea behind Algorithm~\ref{algo:constalgo}, by relaxing the requirement that the largest cluster needs to connect to all the other clusters.
Instead, one can allow more than one level of the tree.
For simplicity, we discuss the case of a matching augmentation ($\delta=1$).
We discuss how to build a tree when $k=\bigO(n^{1-\epsilon})$, for some $0 < \epsilon < 1/2$.
First run the clustering heuristic (line 1), and then sort the clusters by cluster size (number of vertices).
The largest cluster is the root of the tree, and is connected to the second largest cluster, third largest cluster etc., until all vertices in the largest cluster have been shortcut.
Then we repeat the process with the second largest cluster, 
and connect its vertices to the remaining largest clusters, 
placing these clusters one level below.
One might wonder if it is always possible to connect all the clusters into one tree.
For example, if all the clusters are singleton vertices, then this is not possible. 
Or if we have one cluster of four vertices, and five remaining singleton clusters then this is also not possible. 

However, such cases are always avoided.
To see this, we use the fact that any nonnegative integer sequence $d_1, \ldots, d_{k+1} \geq 1$ is the degree sequence of a tree with $k$ edges if and only if $\sum d_i = 2k$.
In our case, every cluster $C_i$ can be seen as a supernode with a maximum degree capacity that is equal to the cluster size $|C_i|$ (every vertex in a cluster can be incident to at most one shortcut edge).
Since $\sum |C_i| = n \geq 2k$, there is in fact always a tree with at most $k$ edges that connects the $k+1$ clusters.
It is also not hard to see that our specific approach (largest clusters first) always works.

The height of this tree will be largest when all the clusters have the same size $\frac{n}{k+1}$, i.e, when branching uniformly.
Assuming $\frac{n}{k+1} \geq 3$, the diameter of the shortcutted graph is at most 
\begin{align*}
(4D^M_{k}+2)\log_{\frac{n}{k+1}-1}(k+1), 
\end{align*}
since it takes at most $2D^M_{k}$ steps to transverse each cluster. If $k = \bigO(n^{1-\epsilon})$ with $0<\epsilon<1/2$, then for $n$ large enough we have $\frac{n}{k+1} \geq 3$, so the diameter is at most $(4D^M_{k}+2)\bigO(\frac{1}{\epsilon})$.

\subsection{Greedy 2-Sweep heuristic}
\label{sec:heuristic}
Lastly, we propose a third intuitive heuristic that works well in practice, but comes without any approximation guarantees. It is inspired by the fast heuristic lower bound called \texttt{2-Sweep} \cite{magnien2009fast}, that is used to estimate the diameter in large graphs.
Algorithm~\ref{algo:g2s} picks a vertex $u$ that is furthest away from a randomly chosen vertex, such that $u$ has less than $\delta$ shortcut edges incident to it. Then it picks a vertex $v$ that is furthest away from $u$, also incident to less than $\delta$ shortcut edges. Add the shortcut edge $\{u,v\}$. Repeat $k$ times.
The \texttt{Greedy 2-Sweep} heuristic attempts to shortcut the two furthest vertices in the graph, with respecting budget constraints. Since determining the two furthest vertices is not feasible for large graphs, this heuristic approximately tries to find the two furthest vertices. 
Running time of this heuristic is $\bigO(k m)$, similar to the previous algorithms.

\begin{algorithm}[t]
\caption{\texttt{Greedy 2-Sweep} heuristic \BCMDd{}.}\label{algo:g2s}
\begin{algorithmic}[1]
\Require graph $G$, $k \geq 1$, $\delta \geq 1$.
\State $G' \leftarrow G$
\State $\forall v\in V: \delta(v) = 0$. 
\MRepeat
\State pick uniform random $u \in \{v \in V: \delta(v) < \delta\}$.
\State $v \leftarrow \text{arg} \max_{\{v \in V: \delta(v) < \delta\}} d_{G'}(u,v)$
\State $G' \leftarrow G' \cup \{u,v\}$.
\State $\delta(v) \leftarrow \delta(v)+1$.
\State $\delta(u) \leftarrow \delta(u)+1$.
\EndRepeat
\end{algorithmic}
\end{algorithm}

\section{Variants}
\label{sec:variants}
We briefly discuss two variants of the \BCMDd{} problem, and show that any approximation algorithm for \BCMDd{} can be used as an approximation algorithm for these variants with a loss of a factor two.

\medskip
\noindent
{\bf Single-source.}
In the single-source variant of \BCMDd{}, we optimize the eccentricity of a designated source vertex instead of the diameter of the entire graph.
\begin{lemma}
\label{lem:ss}
An $\alpha$-approximation for \BCMDd{} is a $2\alpha$-approximation for the single-source variant.
\end{lemma}

\begin{IEEEproof}
The proof is similar to \cite[Section 4]{DemaineShortcut}.
\end{IEEEproof} 

\medskip
\noindent
{\bf Colored diameter.}
Suppose we have a disjoint partition of the vertices $V=V_1 \cup V_2$, where $V_1,V_2 \neq \emptyset$ and $V_1 \cap V_2 = \emptyset$.
Now define the \emph{colored diameter} as the longest shortest path length between a vertex from $V_1$ and a vertex $V_2$.
In several scenarios, it might be the case that one only cares about minimizing the colored diameter instead of the actual diameter.
For example, \cite{Rubenpolarization} used the colored diameter as a measure of polarization between two opposing groups. So in this variant of the \BCMDd{} problem we aim to minimize the colored diameter, by short\-cutting the graph without exceeding a degree budget for each vertex.
It is not hard to see that our algorithms can used for this task as well:

\begin{lemma}
\label{lem:col}
An $\alpha$-approximation for \BCMDd{} is a $2\alpha$-approximation for the colored variant.
\end{lemma}
\begin{IEEEproof}
Let $D^{*}$ (resp. $D_{\text{col}}^{*}$) be the optimum (resp. colored) diameter of the graph after adding $k$ shortcut edges to the graph, while respecting the degree constraints.
Adding $k$ edges that minimize the diameter to $D^{*}$ gives a graph with colored diameter at most $D^{*}$, so we have $D_{\text{col}}^{*}\leq D^{*}$.
On the other hand, adding $k$ edges that minimize the colored diameter to $D_{\text{col}}^{*}$ gives a graph with diameter at most $2D_{\text{col}}^{*}$ (triangle inequality). This implies that $D^{*} \leq 2D_{\text{col}}^{*}$.
So using an $\alpha$-approximation for \BCMDd{}, we get a graph with diameter at most $\alpha D^{*}$.
The colored diameter of this augmented graph is also at most $\alpha D^{*}$, which is at most $2\alpha D_{\text{col}}^{*}$.
\end{IEEEproof}

\section{Inapproximability of \BCMDone{}}
\label{sec:inapprox}

We start our analysis by reducing the \NP-complete \setcover problem \cite{Karp1972} to \BCMDone{} 
to show that we cannot get an approximation ratio better than $4/3$.
The reduction is reminiscent of a reduction proposed by Demaine and Zadimoghaddam~\cite{DemaineShortcut} 
to show an inapproximability result for the single-source version of \BCMD{}, but the analysis is more involved.

\begin{figure}[t]
\centering
\resizebox*{.7\linewidth}{!}{\includegraphics[trim={0.3cm 0.4cm 0.4cm 0},clip,width=1\linewidth]{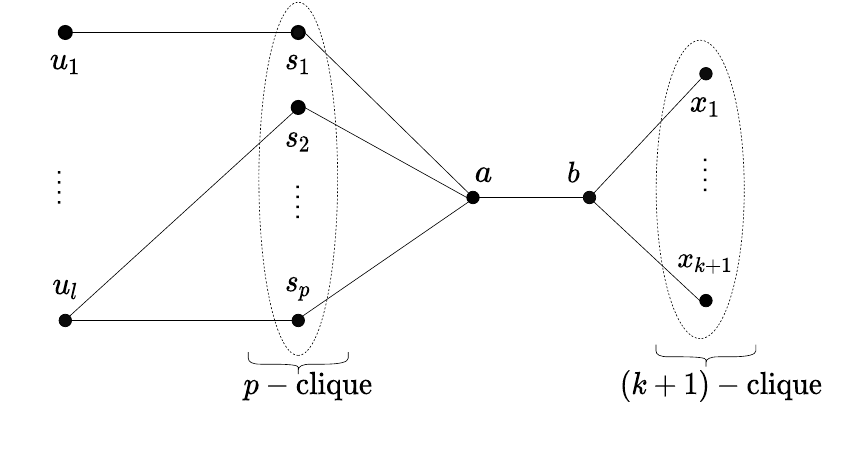}}
\caption{Construction of the graph $H$ used in the reduction from \setcover to \BCMDone{} (Section~\ref{sec:inapprox}).\label{fig:reductionset}}
\end{figure}

An instance of the \setcover problem consists of $p$ sets $S_1, \ldots, S_p$ whose union consists of $l$ items.
The goal is to find the minimum number of sets whose union contains all $l$ items.
We construct a graph $H$ as follows: 
for every set $S_i$ we introduce a vertex $s_i$ and for every item $j$ we introduce a vertex $u_j$.
If an item $j$ is in a set $S_i$ we add an edge from $u_j$ to $s_i$.
All vertices $s_i$ are connected to each other forming an $p$-clique.
We add two more vertices $a$ and $b$. 
Vertex $a$ is connected to all the set vertices $s_i$ and to $b$.
Finally, we add another clique of $k+1$ vertices $x_1, \ldots, x_{k+1}$, and connect them to vertex $b$.
Fig.~\ref{fig:reductionset} shows a schematic example of this construction.

The diameter of graph $H$ is equal to four.
Indeed, the longest shortest paths are between the vertices $u_1, \ldots, u_l$ 
and the vertices $x_1, \ldots, x_{k+1}$, and these are the only shortest paths of length four.
The distance between two vertices $u_i$ and $u_j$ is at most three, 
since we can assume that every item is part of some set 
and all the set vertices $s_1, \ldots, s_p$ form an $p$-clique.
The task of the \BCMDone{} problem is now to add $k$ matching short\-cut edges to $H$ 
such that its diameter is minimized. 
We first prove the following:

\begin{proposition}
\label{prop:red}
There exists a set of $k$ matching short\-cut edges 
that reduce the diameter of $H$ to at most three 
if and only if there exists a set-cover solution with size at most $k$.
\end{proposition}

\begin{IEEEproof}
If there exists a set-cover solution with size at most $k$, 
then we can add at most $k$ matching edges between the vertices $x_1, \ldots, x_k$ 
and the $k$ set vertices that correspond to the set-cover solution.
The exact manner in which they are connected does not matter, 
the resulting graph will have diameter at most three.
Indeed, between any two vertices $u_i$ and $x_j$, there is an edge from $u_i$ to the chosen set which covers $u_i$, then there is a shortcut edge to the $x$-clique, from where we need at most one step to reach $x_j$.

Reversely, we show that if there exists a set of $k$ short\-cut matching edges 
that reduce the diameter to at most three, 
then there exists a set-cover solution with size at most~$k$.
To see this, we prove the existence of a solution that only uses short\-cut edges 
between the $s$-vertices and the $x$-vertices.
We require that $n\geq 2$ and $l,p>k$.
Consider a solution for \BCMDone{} that reduces the diameter of $H$ to three and uses a different short\-cut edge~$e$.
We first examine the easy cases regarding $e$:


\smallskip
\noindent
{\bf (ii)}
Edge $e$ is between some $x_j$ and $a$. 
Because of the matching constraint, any shortest path from a $u$-vertex to $a$ still needs two steps.
Thus, the only purpose of this edge is to reduce the distance between $x_j$ and all the $u$-vertices to three, since any shortest path between $x_{j'}$ ($j' \neq j$) and some $u_i$ that uses this edge still has length four.
We can replace this edge by an edge between $x_j$ and any set vertex, the distance between $x_j$ and all the $u$-vertices is still at most three.

\smallskip
\noindent
{\bf (iii)}
Edge $e$ is between some $u_i$ and $a$. 
The only purpose of this edge is to reduce the distance between $u_i$ and the $x$-vertices to three.
One can replace this edge with an edge between a set vertex that is connected to $u_i$ and some $x_j$. The distance between $u_i$ and the $x$-vertices is still at most three.

\smallskip
\noindent
{\bf (iv)}
Edge $e$ is between some $u_i$ and $b$. The only purpose of this edge is to reduce the distance between $u_i$ and all the $x_j$ to two.
One can replace this edge with an edge between a set vertex that is connected to $u_i$ and some $x_j$. The distance between $u_i$ and the $x$-vertices is still at most three.

\smallskip
\noindent
{\bf (v)}
Edge $e$ is between some $s_i$ and $b$. The only purpose of this edge is to reduce the distance between the item vertices connected to $s_i$ and all the $x_j$ to three.
One can replace this edge with an edge between $s_i$ and some $x_j$. The distance between all the item vertices connected to $s_i$ and the $x$-vertices is still at most three.

\smallskip
\noindent
{\bf (vi)}
Edge $e$ is between two $u$-vertices. The only possible purpose of this edge would be to reduce the distance between one of the two $u$ vertices and one specific $x_i$ vertex to three, since both endpoints of $e$ have already been shortcut. Replacing this edge by an edge between a corresponding set vertex and any $x$ vertex results in a graph where these shortest paths are still at most three.

\smallskip
\noindent
{\bf (vii)}
Edge $e$ is between some $u_i$ and $s_j$. The only purpose of this edge would be to reduce the distance between $u_i$ and one specific $x_k$ to three (by shortcutting $a$), since $s_j$ has already been shortcut.
Replacing this edge by an edge between a $s_j$ and any $x$ vertex results in a graph where these shortest paths are still at most three.

After rewiring edges according to the previous cases, we end up with a solution where all of the shortcut edges are either between a $u$-vertex and an $x$-vertex, or an $s$-vertex and an $x$-vertex.
Suppose a vertex $u_i$ does not have a shortcut edge to any of the $x$-vertices.
We can prove that there must exist a set vertex that covers $u_i$ and is shortcut to one of the $x$-vertices.

Suppose, by contradiction, that $u_i$ is not shortcut to the $x$-vertices, and does not have a covering set vertex that is shortcut to the $x$-vertices. 
There are two cases. First, consider the case when $u_i$ has a shortest path that first visits another $u_j$ vertex, and then a shortcut edge from $u_j$ to one of the $x$-vertices. This would take at least three steps, since it takes at least two steps to move from $u_i$ to $u_j$. Moreover, it would only reduce the distance from $u_i$ to one specific $x$-vertex to three (since we need an extra step to reach the other $x$-vertices). Secondly, suppose $u_i$ follows a shortest path that first visits a set vertex containing $u_i$, and then to another set vertex (which does not contain $u_i$) that is shortcut to the $x$-vertices.  This also requires three steps, and again it would only reduce the distance from $u_i$ to one specific $x$-vertex to three.
Since there are $k+1$ $x$-vertices, but only $k$ shortcut edges which reduce the distance to three for only one specific $x$-vertex, there is always one $x$-vertex for which the distance to $u_i$ is four, which contradicts the optimality.

Hence, every $u_i$ that is not shortcut to the $x$-vertices is connected to some set vertex that is
shortcut to the $x$-vertices. Additionally, if $u_i$ is shortcut to an $x_j$ vertex, we can replace that edge by an edge between the corresponding set vertex and $x_j$. So we have constructed a set-cover of size at most $k$.
\end{IEEEproof}

We are ready to prove the inapproximability result for \BCMDone{}. 
Clearly the result also holds for the \BCMDd{} problem, as it includes \BCMDone{} as a special instance.

\begin{lemma}
\label{lem:inapprox}
There exists no polynomial-time $(\frac{4}{3}-\epsilon)$-approximation algorithm 
for the \BCMDone{} problem, 
assuming $\poly \neq \NP$.
\end{lemma}

\begin{IEEEproof}
Assume that there exists a $(\frac{4}{3}-\epsilon)$-approximation algorithm $A$ for \BCMDone{}.
We can use algorithm $A$ to decide \setcover, 
i.e., decide if there exists a set cover of size at most $k$ to a \setcover instance.
If there is a set cover of size at most $k$, 
according to Proposition~\ref{prop:red} the optimum of \BCMDone{} in graph $H$ is at most three.
Hence, algorithm $A$ will return a solution with cost at most $(\frac{4}{3}-\epsilon) 3 < 4$. 
Because the diameter only takes integer values, the diameter of the augmented graph is at most three. 
On the other hand, if there does not exist a set cover of size at most $k$, 
according to Proposition~\ref{prop:red} again there is no $k$-augmentation 
that reduces the diameter of $H$, and thus, algorithm $A$ will not reduce the diameter. 
So we can use algorithm $A$ to decide the \setcover problem, 
and 
the solution returned by algorithm $A$ can be used to recover a solution for \setcover,
in case of a \texttt{yes}-instance.
\end{IEEEproof}

\section{Experimental evaluation}
\label{sec:exps}
In this section we compare and evaluate the practical performance of our three main algorithms from Section~\ref{sec:gengraphs}.
For our first two algorithms, the initial center vertex (resp., segment) is chosen randomly.
We use five real-life networks, shown in Table~\ref{tab:datasets},  
of varying types, and all with a reasonably large diameter of their largest connected component.

Since computing the diameter is not feasible for large graphs ($\bigO(n m)$ time using BFS or $\bigO(n^{2.372}\cdot \text{polylog}(n))$ time using matrix multiplication), we will use the fast heuristic lower bound called \texttt{2-Sweep} \cite{magnien2009fast}, which has been reported to work very accurately for numerous real-life networks \cite{Pierluigi,magnien2009fast}. The main idea is to pick the farthest vertex from a random vertex, and return its eccentricity. This heuristic lower bound needs $\bigO(m)$ time for connected graphs.
The main goal is to compare our three algorithms against each other, but we also compare with a straightforward baseline: uniformly at random. This might seem simple, but it has been reported that random augmentations often perform quite well for diameter reduction problems \cite{BollobasRandom}. 
The random algorithm selects uniformly at random $k$ shortcut non-edges, 
while respecting a degree increase of at most $\delta$ per vertex. 
All experiments are performed on an Intel\,core\,i5 machine at~1.8 GHz with 16\,GB\,RAM. 
All methods are implemented in Python~3.8 and are publicly available.\footnote{https://www.dropbox.com/sh/w9h1g2js54rjiy4/AADSsJwlOicUgC2tVk\newline gdr7Rya?dl=0}

\begin{figure*}[!htbp]
\centering
\subfloat[Brightkite $(\delta=1)$\label{fig:toya}]{\scalebox{1}{\includegraphics[width=0.6\columnwidth]{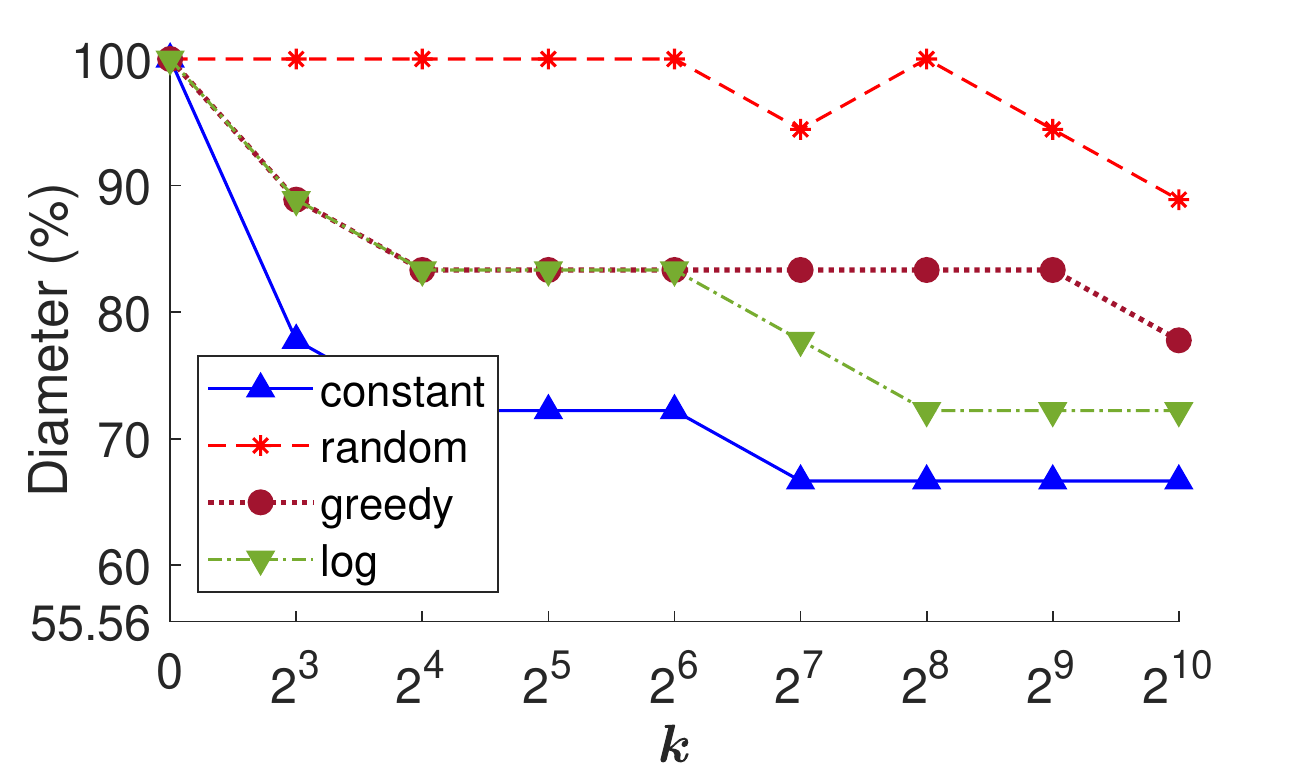}}}
\subfloat[Brightkite $(\delta=25)$\label{fig:toyb}]{\scalebox{1}{\includegraphics[width=0.6\columnwidth]{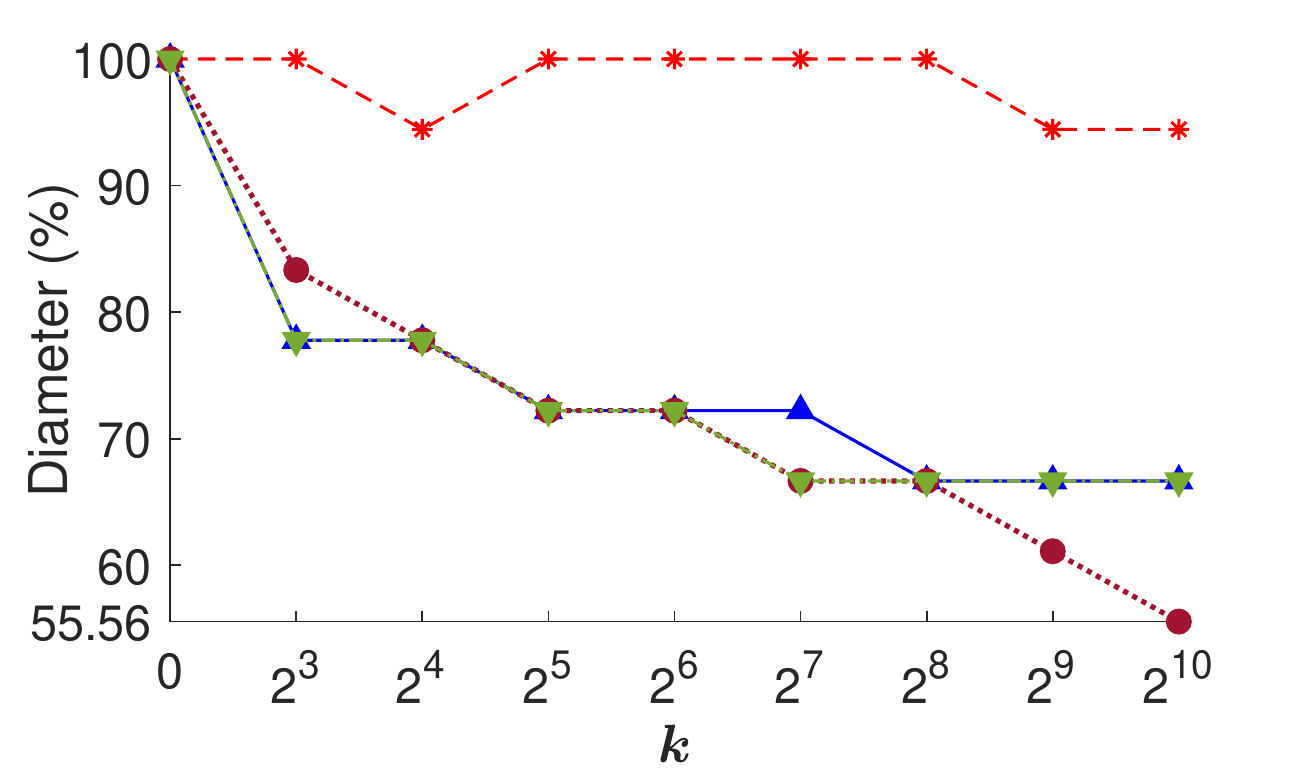}}}
\subfloat[Brightkite $(\delta=2^{10})$\label{fig:toyc}]{\scalebox{1}{\includegraphics[width=0.6\columnwidth]{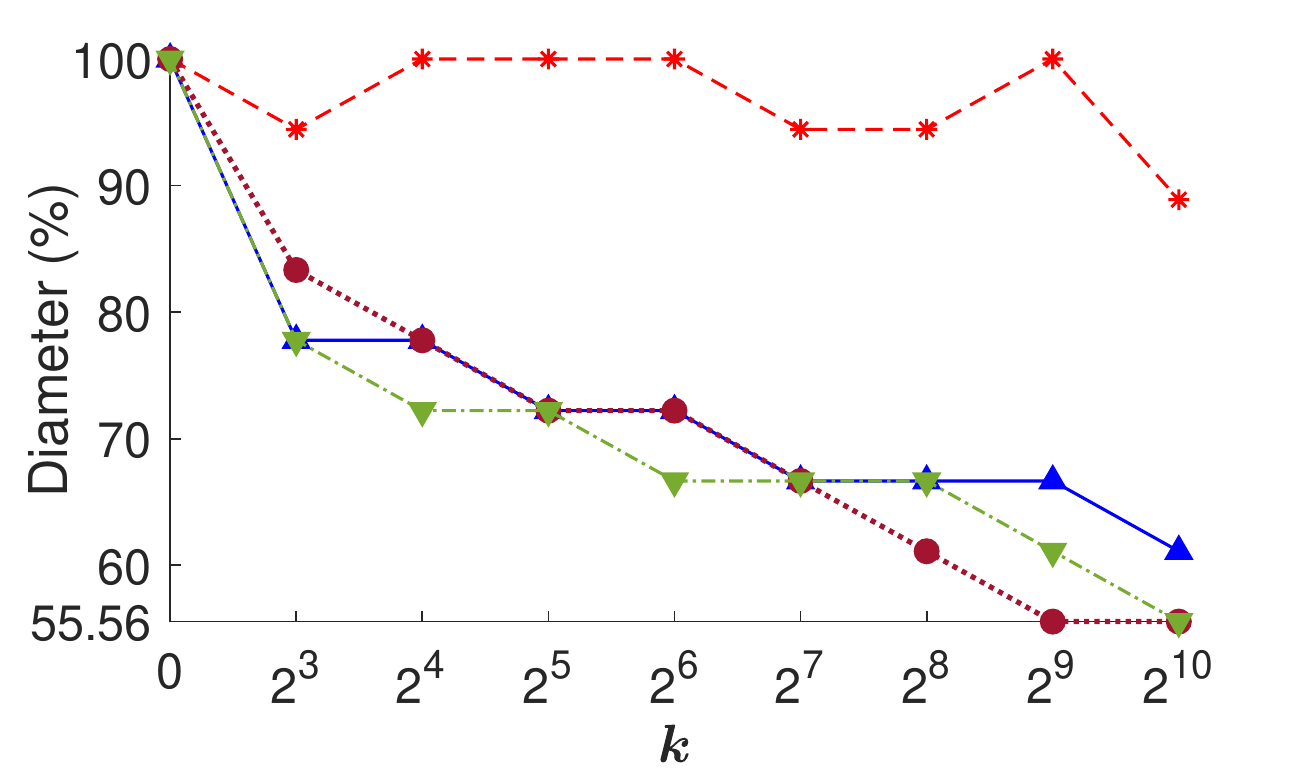}}}
\\
\subfloat[Reactome $(\delta=1)$\label{fig:toya}]{\scalebox{1}{\includegraphics[width=0.6\columnwidth]{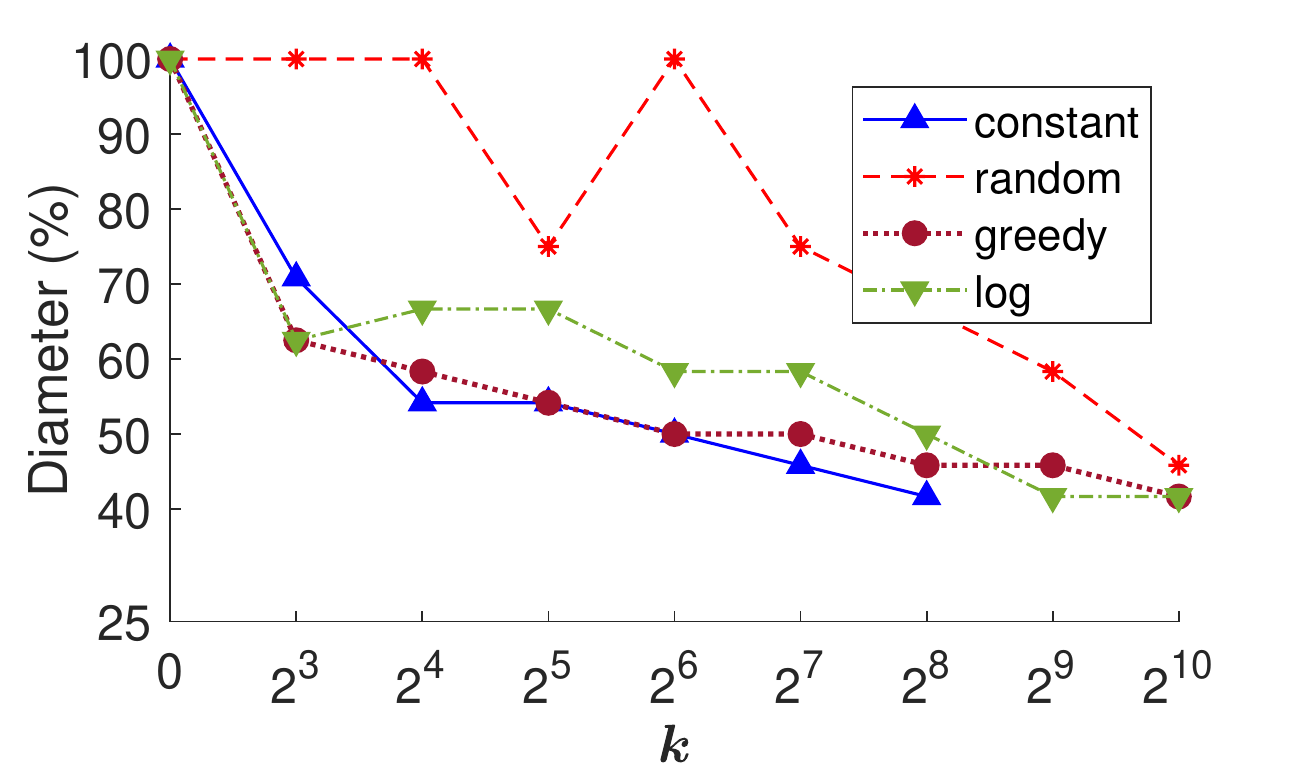}}}
\subfloat[Reactome $(\delta=25)$\label{fig:toyb}]{\scalebox{1}{\includegraphics[width=0.6\columnwidth]{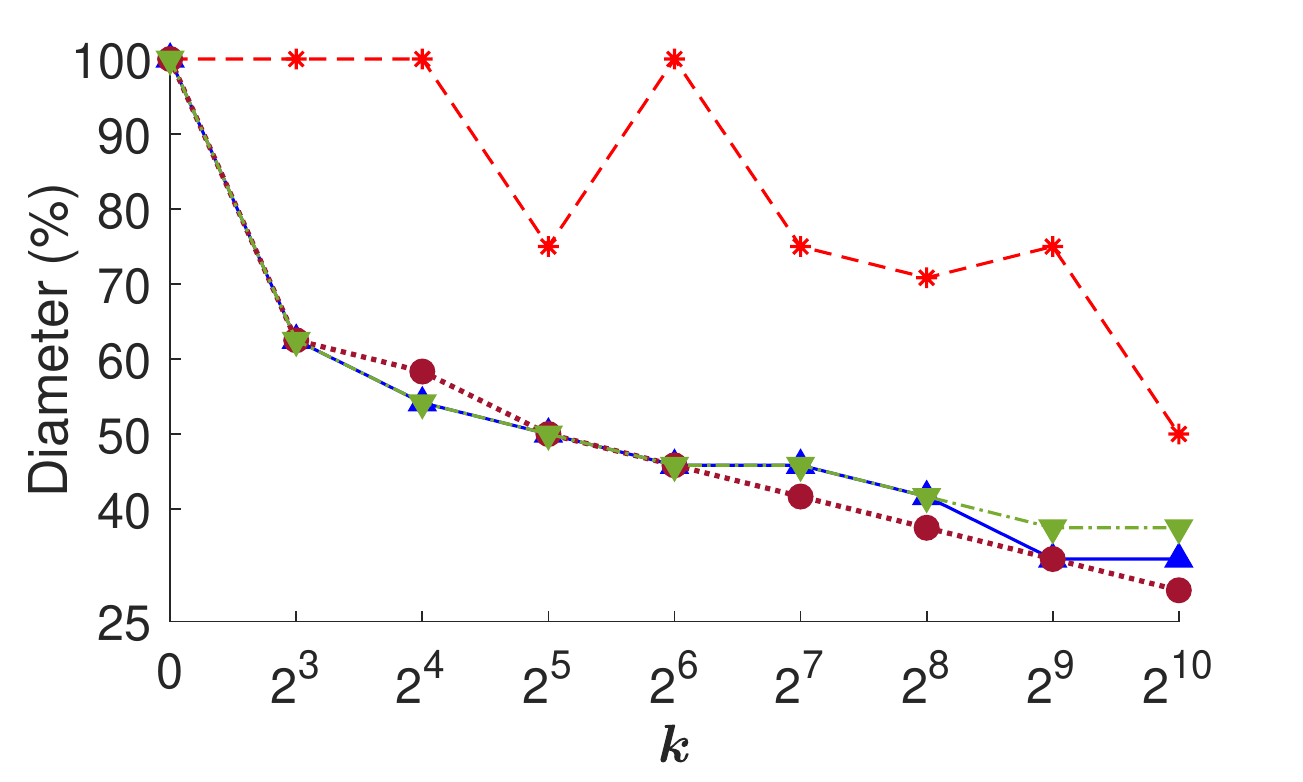}}}
\subfloat[Reactome $(\delta=2^{10})$\label{fig:toyc}]{\scalebox{1}{\includegraphics[width=0.6\columnwidth]{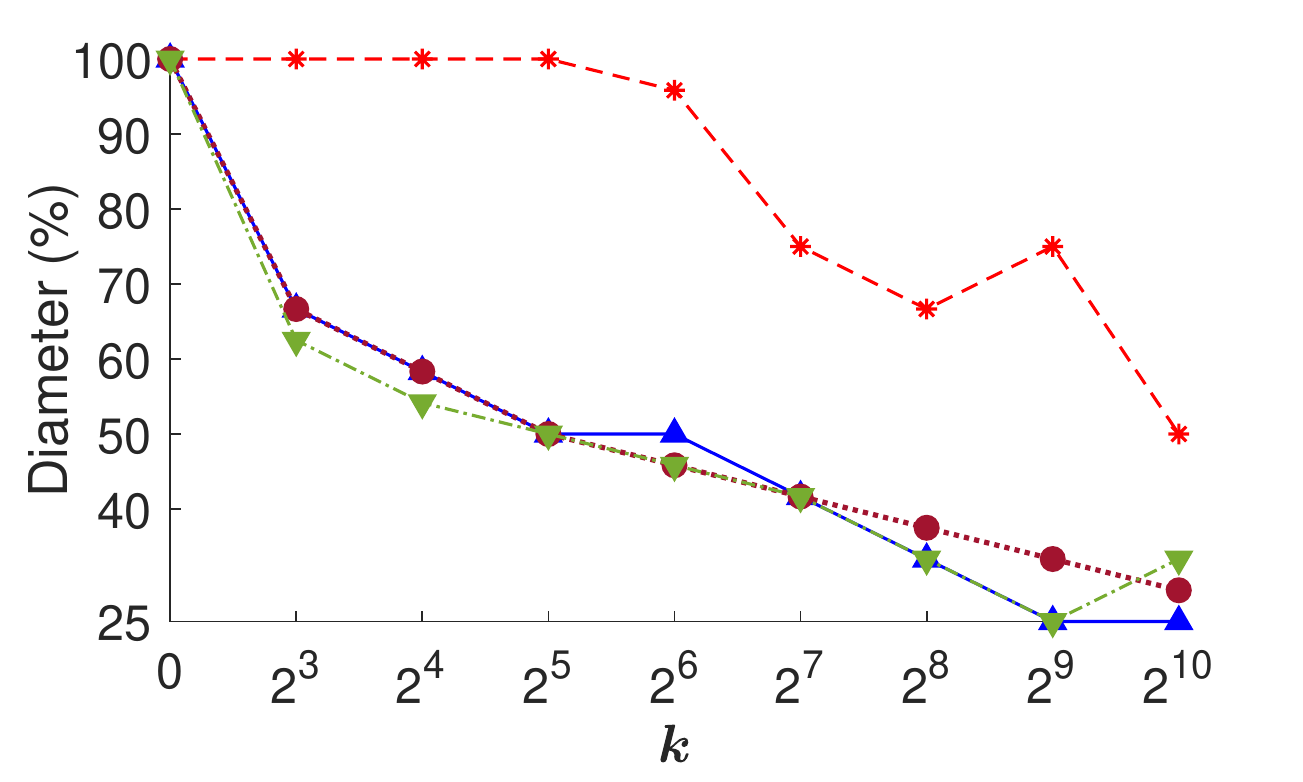}}}
\\
\subfloat[Power $(\delta=1)$]{\scalebox{1}{\includegraphics[width=0.6\columnwidth]{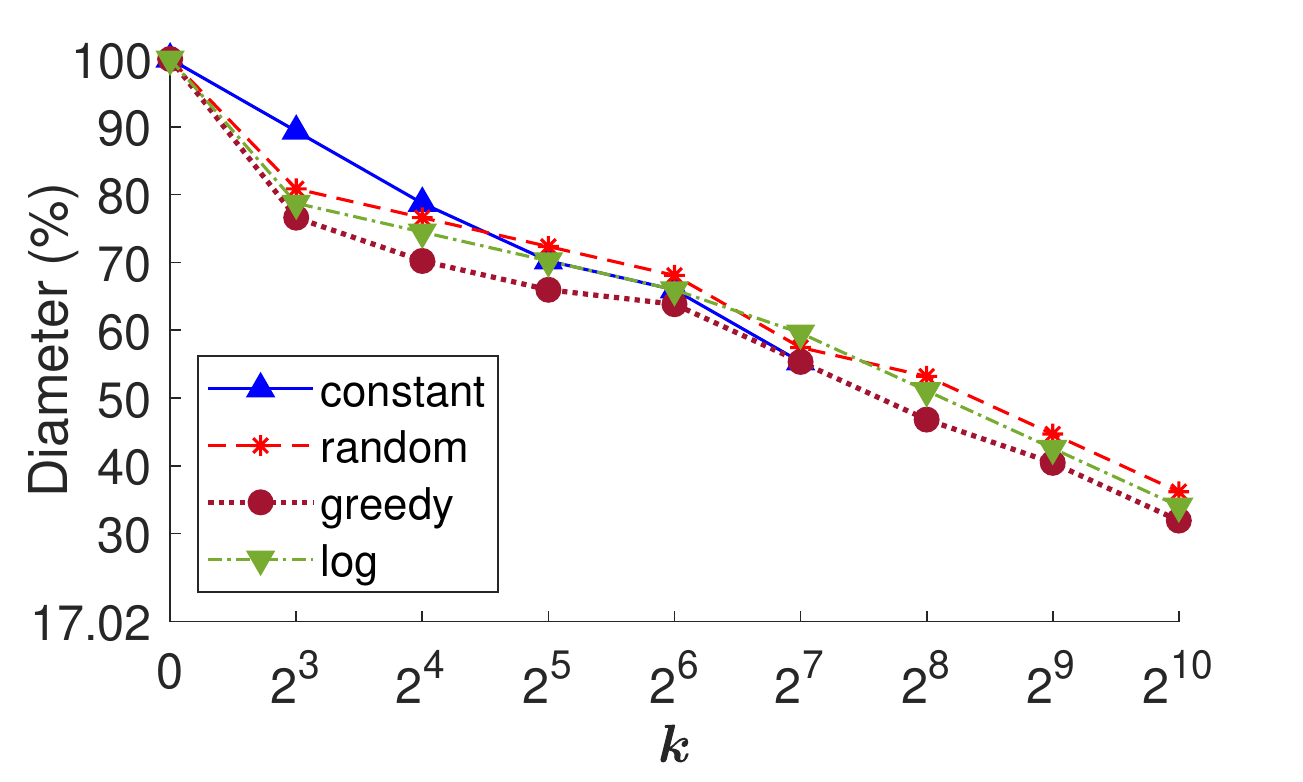}}}
\subfloat[Power $(\delta=25)$]{\scalebox{1}{\includegraphics[width=0.6\columnwidth]{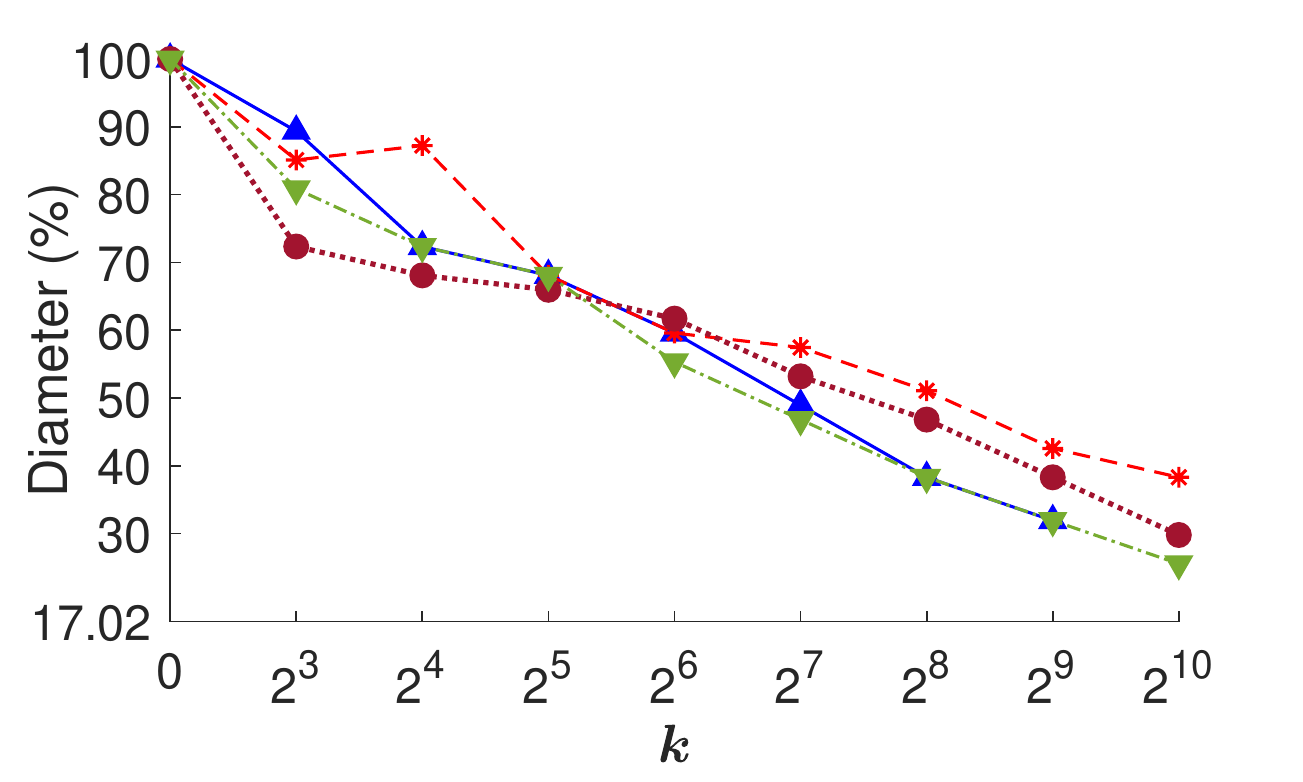}}}
\subfloat[Power $(\delta=2^{10})$]{\scalebox{1}{\includegraphics[width=0.6\columnwidth]{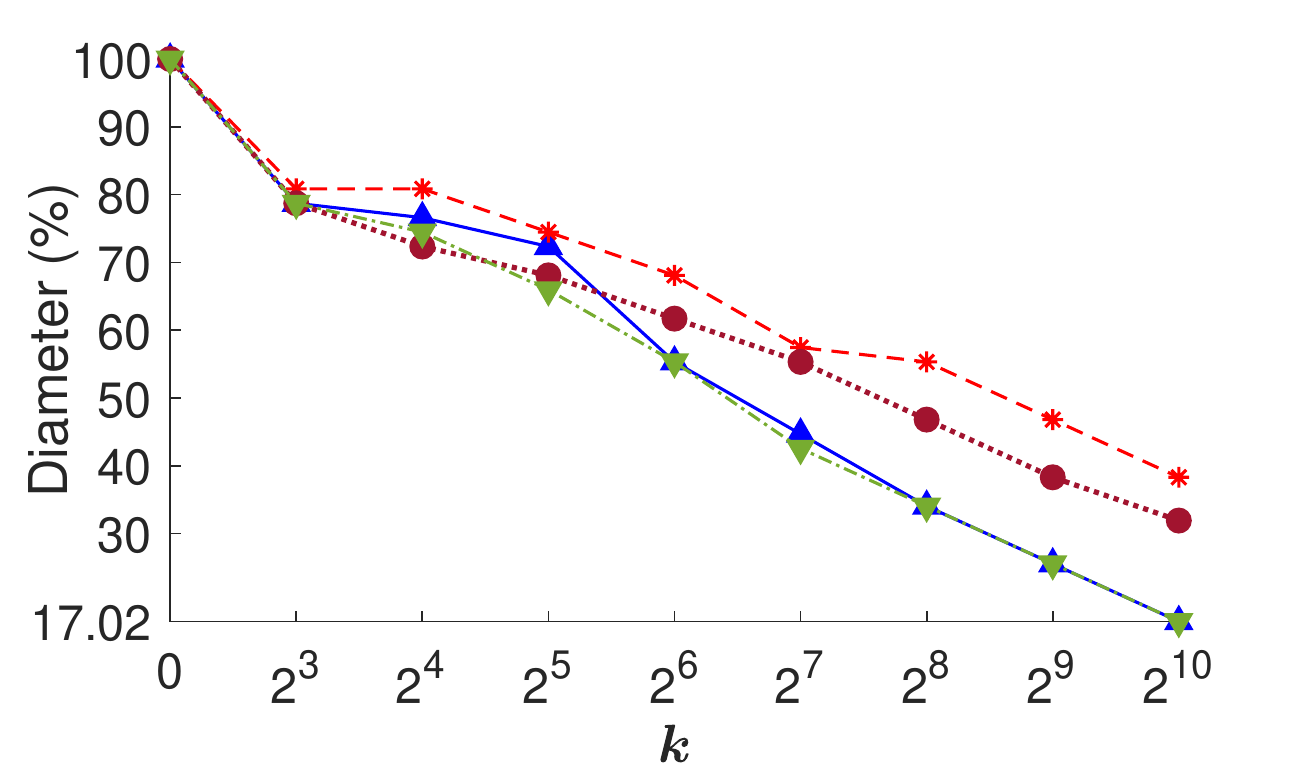}}}
\\
\subfloat[Amazon $(\delta=1)$]{\scalebox{1}{\includegraphics[width=0.6\columnwidth]{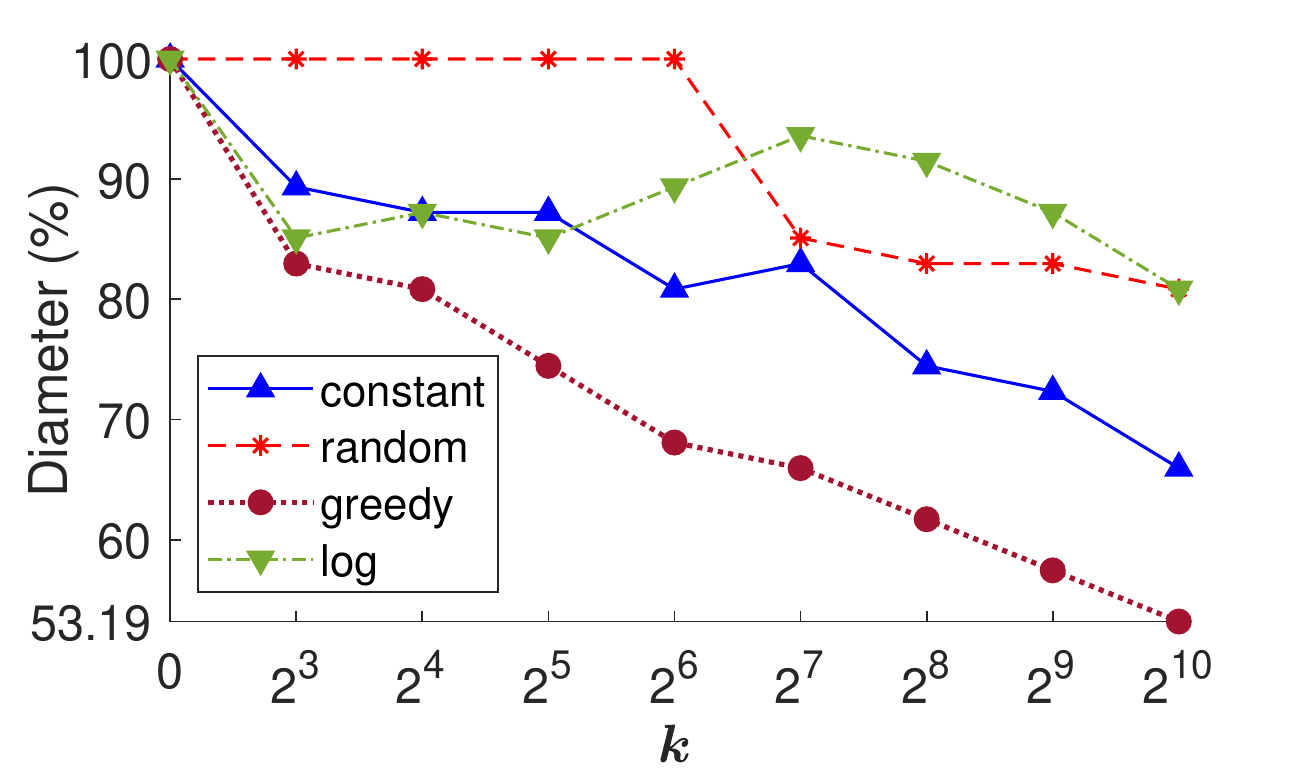}}}
\subfloat[Amazon $(\delta=25)$]{\scalebox{1}{\includegraphics[width=0.6\columnwidth]{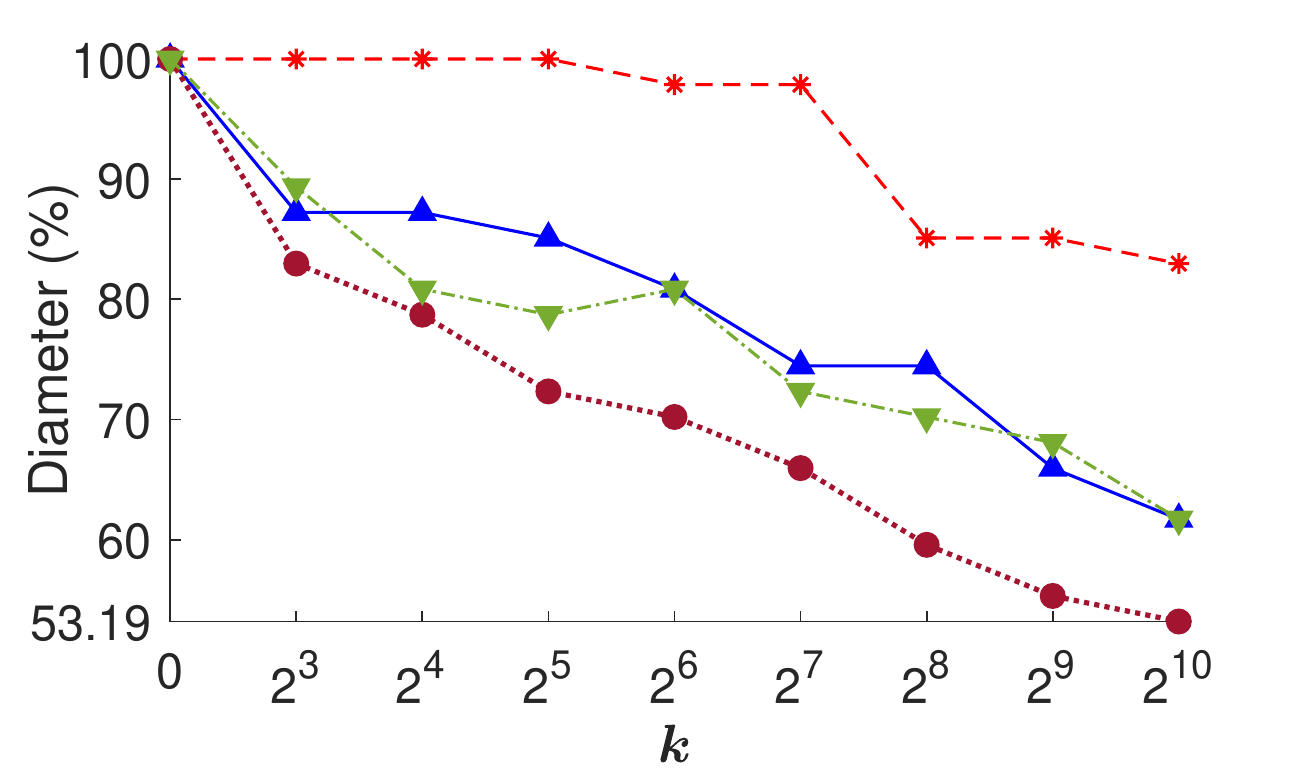}}}
\subfloat[Amazon $(\delta=2^{10})$]{\scalebox{1}{\includegraphics[width=0.6\columnwidth]{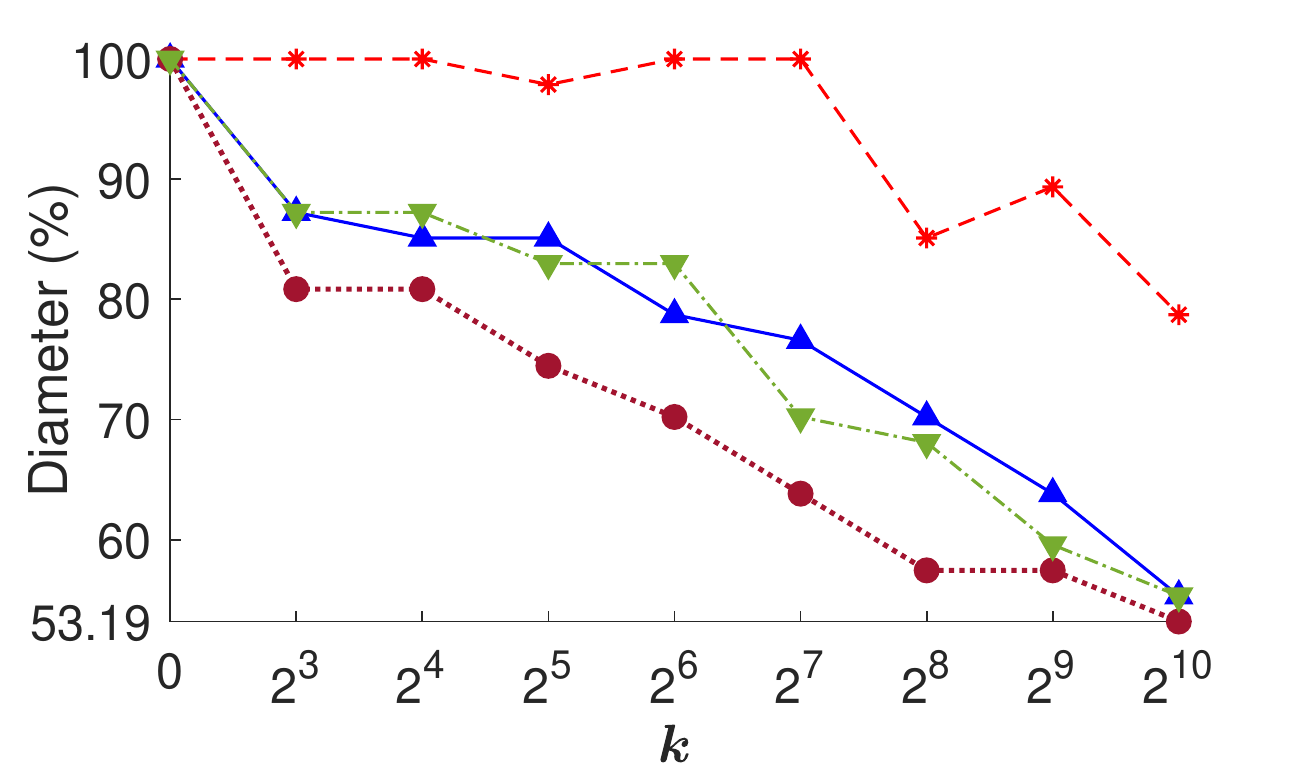}}}
\\
\subfloat[roadNet-PA $(\delta=1)$]{\scalebox{1}{\includegraphics[width=0.6\columnwidth]{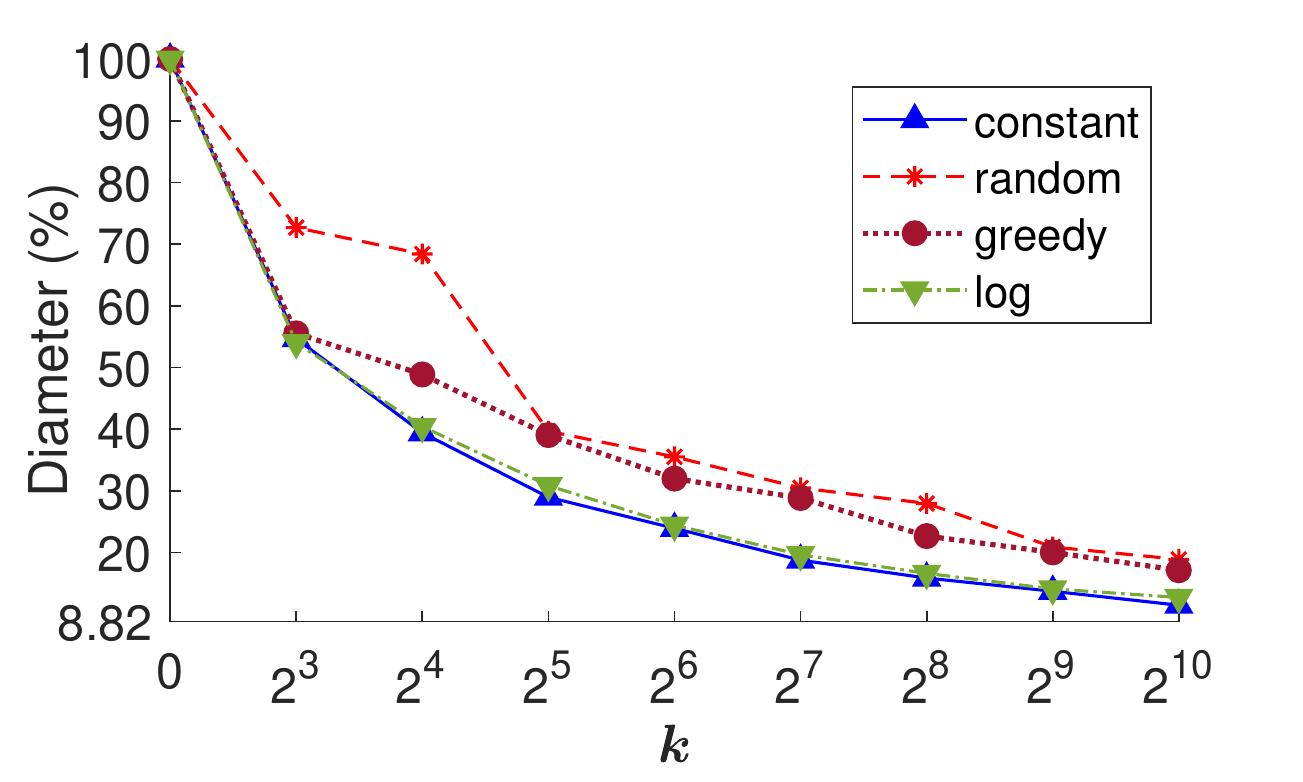}}}
\subfloat[roadNet-PA $(\delta=25)$]{\scalebox{1}{\includegraphics[width=0.6\columnwidth]{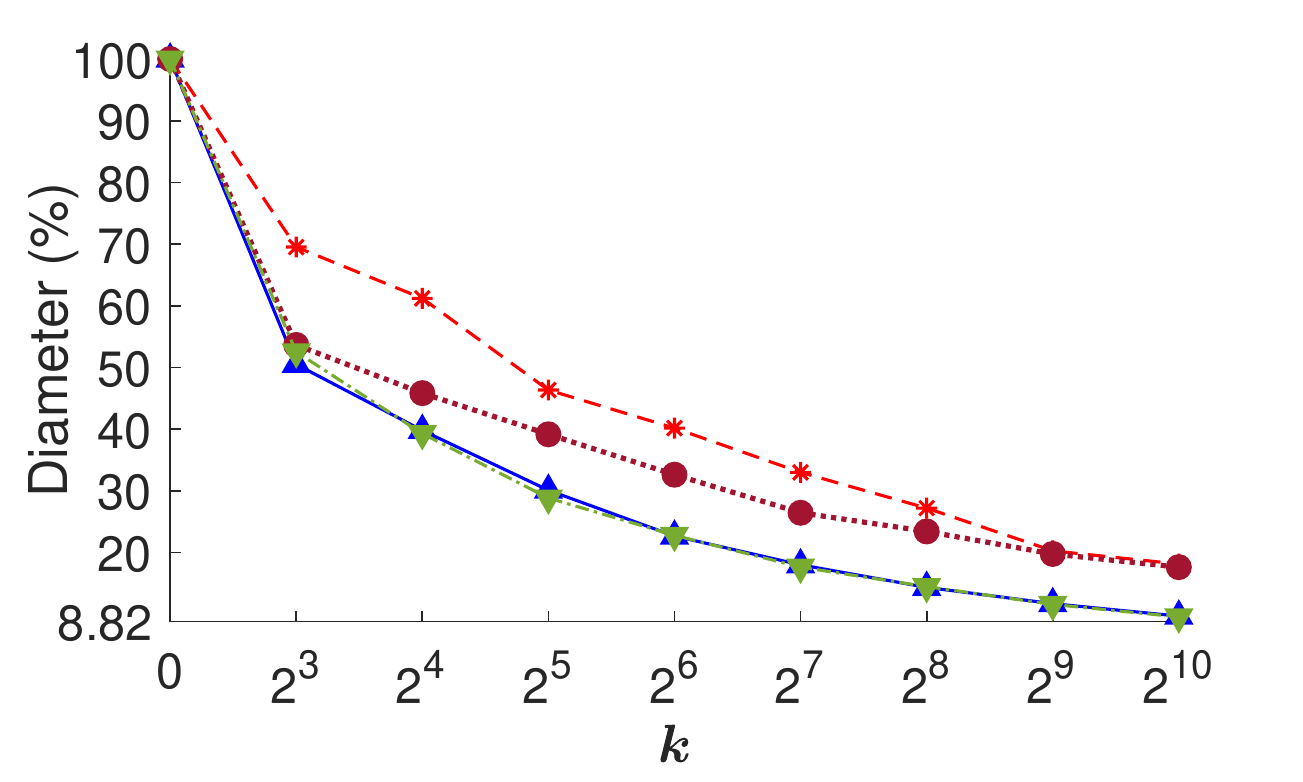}}}
\subfloat[roadNet-PA $(\delta=2^{10})$]{\scalebox{1}{\includegraphics[width=0.6\columnwidth]{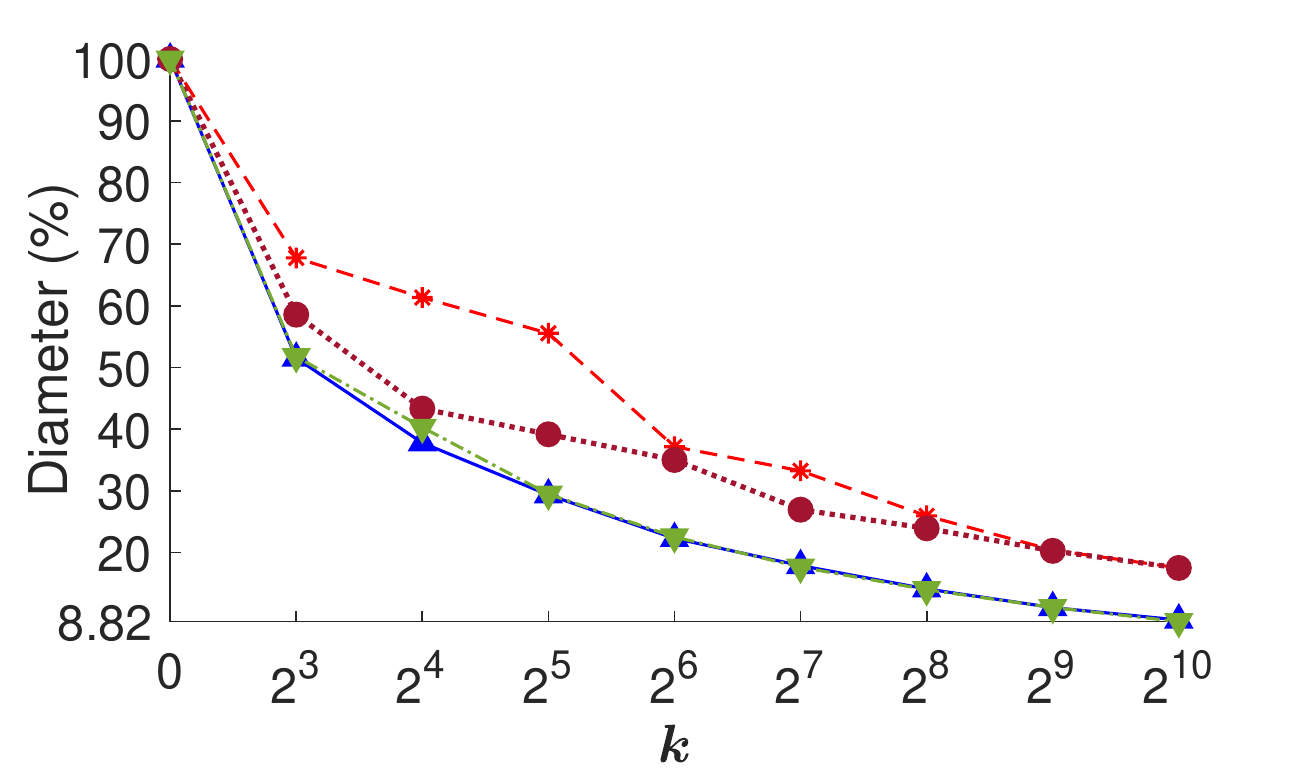}}}
\caption{The performance of the proposed algorithms for various settings of $k$ and $\delta$.
The interruption of the blue colored plot (constant-factor approximation algorithm) indicates that the criterion for the algorithm to run is not fulfilled, as discussed in Section~\ref{sec:smallk}.\label{fig:expk}}
\end{figure*}

\begin{table}
  \renewcommand{\arraystretch}{1.3}
  \caption{Overview of the real-life datasets used in the experiments. For each network, we extracted the largest connected component (LCC). $D$ denotes the diameter.\label{tab:datasets}}
  
  \footnotesize
  \begin{tabular}{llrrr}
    \toprule
    Data (LCC) & Type & $|V|$ & $|E|$  & $D$ \\
    \midrule
    Brightkite \cite{Konect,MovementLesko} & social & 56\,739 & 212\,945 & 18 \\
    Reactome \cite{Konect,Reactome} & protein& 5\,973 & 145\,778 & 24 \\
    Power grid \cite{Konect,watts1998collective} & infrastructure & 4\,941 & 6\,594 & 46 \\
    Amazon \cite{snapnets,yang2012defining} & co-purchasing & 334\,863 & 925\,872 & 47\\
    roadNet-PA \cite{snapnets,leskovec2008community} & road & 1\,088\,092 & 1\,541\,898 & 794\\		
  \bottomrule
\end{tabular}
\end{table}

\begin{figure}[t]
\centering
\subfloat[roadNet-PA\label{fig:ta}]{\scalebox{1}{\includegraphics[width=0.5\columnwidth]{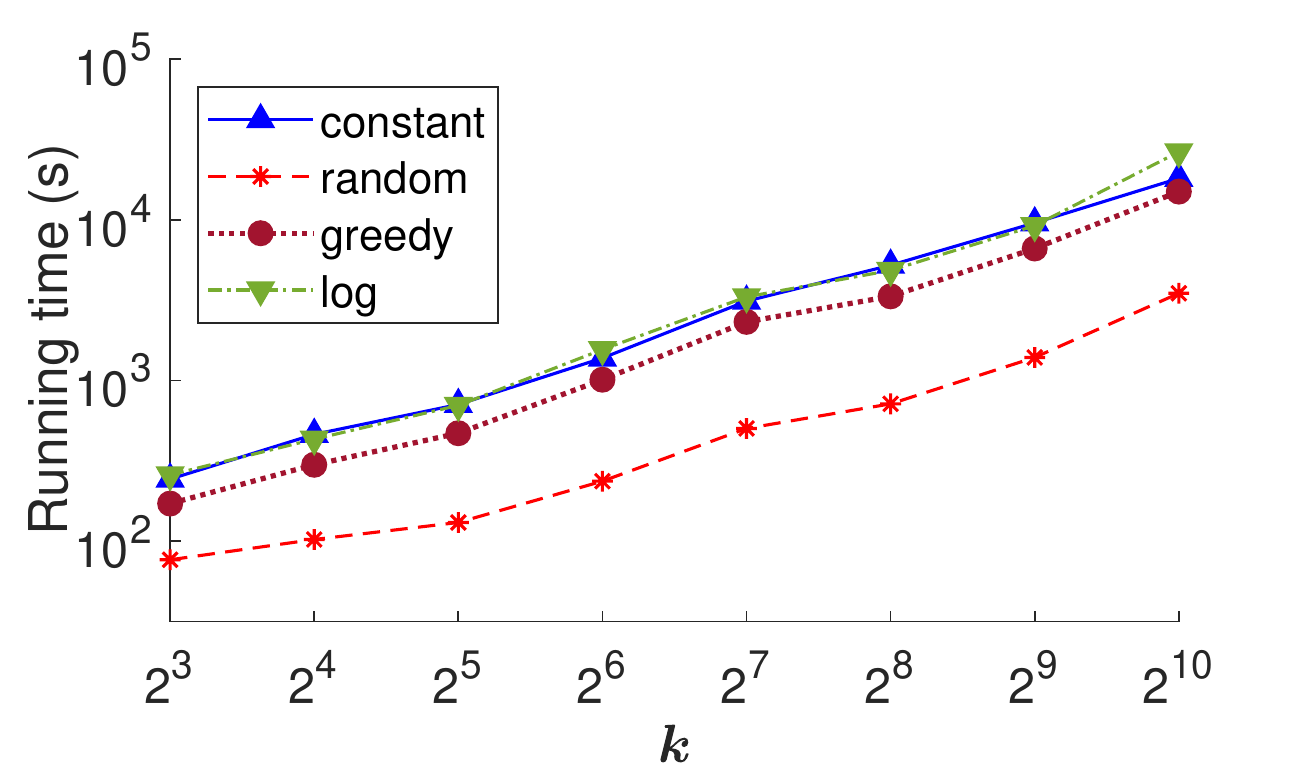}}}
\subfloat[Amazon\label{fig:tb}]{\scalebox{1}{\includegraphics[width=0.5\columnwidth]
{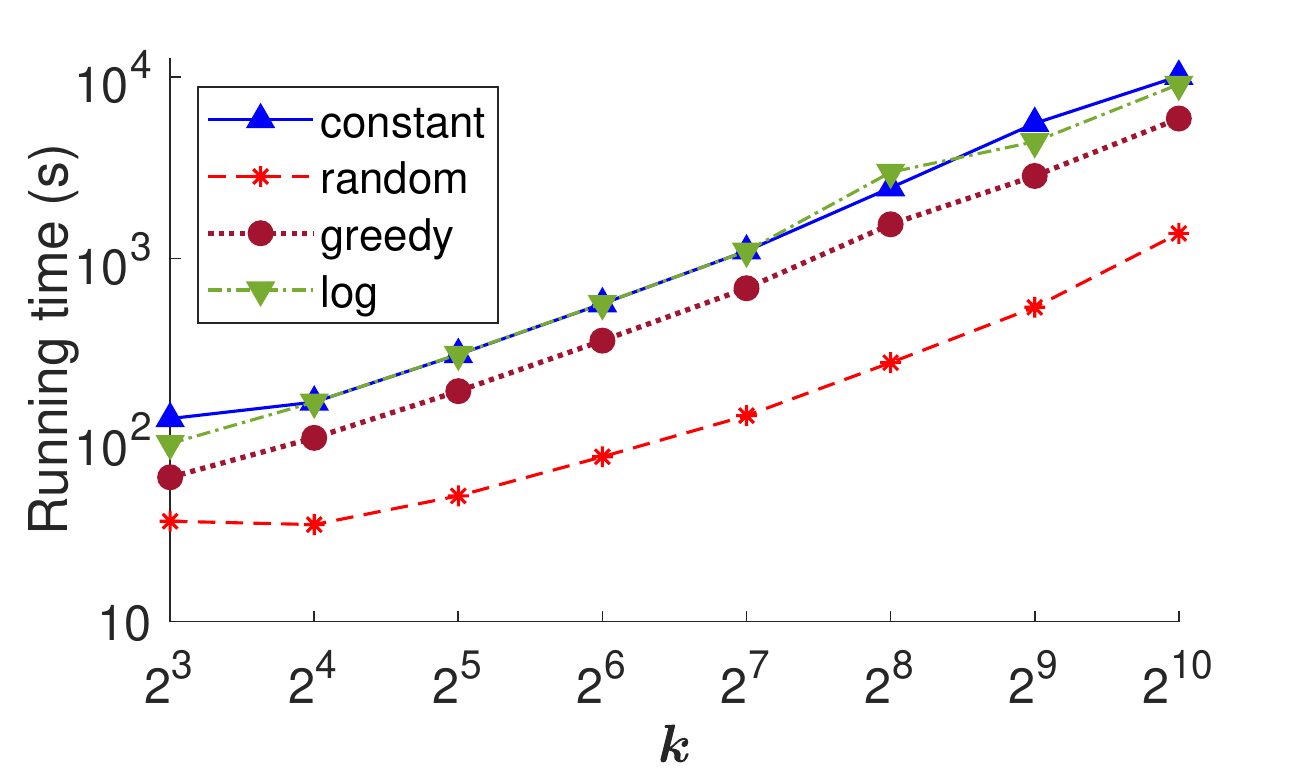}}}
\caption{Running time as a function of the number of shortcut edges $k$, on the Amazon and roadNet-PA datasets.\label{fig:runningtime}}
\end{figure}

\smallskip
\noindent
{\bf Influence of $k$ and $\delta$.}
We test the influence of the number of shortcut edges $k$ and the degree budget parameter $\delta$ 
on the performance of the proposed algorithms.
We let $k \in \{8,16,...,1024\}$ and 
$\delta \in \{1,25,1024\}$, 
where $\delta=1024$ corresponds to an unlimited degree increase budget (given our range for $k$). 
For every $k$ and $\delta$, we repeat each algorithm five times and return the minimum diameter. 
Results are shown in Fig.~\ref{fig:expk}. 
The Greedy \texttt{2-Sweep} heuristic, the log-approximation 
and the constant-factor approximation 
all significantly perform better than randomly adding edges. 
Interestingly, for the Power and roadNet-PA datasets adding random edges can work well to reduce the diameter, 
even with a matching constraint. 
Another interesting observation is that one does not need a high degree budget $\delta$ to obtain good diameter reduction. Even with matching shortcut edges ($\delta=1$), 
a decent reduction in diameter is often possible.
The difference in results between $\delta=25$ and $\delta=1024$
does not seem to be that large, 
but is more noticeable for larger values of $k$. 
The interruption of the blue colored plot means 
that the constant-factor approximation does not meet the criteria to run. 
This happens whenever the largest cluster does not have enough vertices (with enough budget) 
to shortcut to the remaining clusters. 
In Section~\ref{sec:smallk} we showed that the algorithm is guaranteed to work if $k \leq \sqrt{\delta n}-1$, 
but in practice the algorithm did run for larger values of $k$.

\smallskip
\noindent
{\bf Running time.} 
Fig.~\ref{fig:runningtime} shows the running time as a function of $k$ on the two largest datasets.
Doubling the number of edges $k$ roughly leads to doubling the running time. 
This corresponds to the $\bigO(k m)$ time complexity of
our methods and the baseline.
The influence of $\delta$ on the running time is~negligible.

\section{Conclusion and future work}

We study the problem of adding $k$ new edges to a graph 
in order to minimize its diameter
and while respecting degree constraints on the number of edges added at each vertex.
We present three algorithms for this task, and show the hardness of approximating the problem with ratio better than 4/3. Our paper opens up several directions for future work.
First, an important challenge is to close the approximability gap between lower and upper bound.
A problem variant, motivated by real-world applications, 
is to consider non-unit costs for the new edges to be added. 

\section*{Acknowledgment}

Both authors are supported by the ERC Advanced Grant REBOUND (834862),
the EC H2020 RIA project SoBigData (871042), and the Wallenberg AI, Autonomous Systems
and Software Program (WASP) funded by the Knut and Alice Wallenberg Foundation

\bibliographystyle{IEEEtran}
\bibliography{IEEEabrv,icdm-paper-diameter}
\end{document}